# DESIGNING ZONAL-BASED FLEXIBLE BUS SERVICES UNDER STOCHASTIC DEMAND

Enoch LEE [a], Xuekai CEN* [b], Hong K. LO [a], and Ka Fai Ng [a]
[a] Department of Civil and Environmental Engineering,
The Hong Kong University of Science and Technology, Hong Kong, China
[b] School of Traffic & Transportation Engineering, Central South University, China
* Corresponding author
Email addresses: eleead@ust.hk (Enoch LEE), cehklo@ust.hk (Hong K. LO),
cenxuekai@csu.edu.cn (Xuekai CEN), kfngaa@connect.ust.hk (Ka Fai Ng)

**ABSTRACT**

In this paper, we develop a zonal-based flexible bus services (ZBFBS) by considering both passenger demands' spatial (origin-destination or OD) and volume stochastic variations. Service requests are grouped by zonal OD pairs and number of passengers per request, and aggregated into demand categories which follow certain probability distributions. A two-stage stochastic program is formulated to minimize the expected operating cost of ZBFBS, in which the zonal visit sequences of vehicles are determined in Stage-1, whereas in Stage-2, service requests are assigned to either regular routes determined in Stage-1 or ad hoc services that incur additional costs. Demand volume reliability and detour time reliability are introduced to ensure quality of the services and separate the problem into two phases for efficient solutions. In phase-1, given the reliability requirements, we minimize the cost of operating the regular services. In phase-2, we optimize the passenger assignment to vehicles to minimize the expected ad hoc service cost. The reliabilities are then optimized by a gradient-based approach to minimize the sum of the regular service operating cost and expected ad hoc service cost. We conduct numerical studies on vehicle capacity, detour time limit and demand volume to demonstrate the potential of ZBFBS, and apply the model to Chengdu, China, based on real data to illustrate its applicability.

Keywords: flexible bus, demand responsive transit, stochastic programming, reliability

## 1. INTRODUCTION

Public transport (PT), if successful, is instrumental in addressing the transportation needs of many metropolitan areas. Traditionally, without real-time demand information, PT services are operated on fixed routes with fixed schedules (FRFS), regardless of the actual demand, which are neither efficient nor cost-effective. As for demand responsive services, such as taxi, while flexible, they are too costly to serve as a major PT mode due to lack of economies of scale. With the emergence of information technology, travelers can request for ride services in real-time. Transportation network companies (TNCs), such as Lyft, Uber, and Didi Chuxing, can match travelers to vehicles almost instantaneously to provide for ride services, typically through small vehicles that give high flexibility but may worsen traffic congestion in metropolitan cities (Schaller, 2018). TNCs open up the possibility of demand responsive ride sharing services that can be operated at lower costs due to economies of scale, and yet yield much more flexibility than FRFS services, e.g., Didi Bus. The key is to devise a system to allow passengers sharing the same vehicles, which will achieve economies of scale, and at the same time restrict the detour time to pick up multiple passengers to an acceptable level.

The concept of demand responsive ride sharing services is not new, which is also referred to as dial-a-ride (DAR) services where customers dial calls for service requests, and the associated problem is named as the dial-a-ride problem (DARP). Such DAR services are available today but limited in scopes, such as airport trips, or trips for the elderly, people with disabilities or special needs. What is new now is that such service requests, enabled by service platforms such as Uber and Didi, are gaining popularity to the extent that demand responsive transit (DRT) may become a substantive PT mode, especially in view of the widely recognized Mohring effect: increasing passenger volumes allows for improved services, initiating a virtuous cycle further attracting new users. In fact, such demand responsive services are gaining recognition in China, referred to as customized bus service, serving people between central business area and residential area especially in peak hours (Liu and Ceder, 2015).

DARPs are mainly about designing vehicle routes and schedules for service requests between origins and destinations, and can be considered as a generalization of the pick-up and delivery vehicle routing problem (PDVRP) and the vehicle routing problem with time windows (VRPTW) (Cordeau and Laporte, 2007). Important features of DARPs include vehicle capacity and time window or detour time for picking up and dropping off passengers in a reasonable limit. Ho et al. (2018) divided DARPs into deterministic and stochastic DARPs, which depend on whether the information (such as demand) is known with certainty or not, and stated that few studies focused on static and stochastic DARPs. In this paper, we aim to develop a novel zonal-based flexible bus system, which is a demand responsive transit service considering stochasticity of zonal demand volume and detour time within each zone, vehicle capacity, and detour constraint for each bus.

Flexible bus is a class of DRT that provides door-to-door service with a relatively low price, which can bring a higher social welfare to an area with high population density (Zhang et al., 2018). Some researchers studied DRT with analytical models, which typically assumed that the demand is uniformly distributed over space (e.g., Chen and Nie, 2017a, 2017b; Diana et al., 2006; Kim and Schonfeld, 2014, 2015; Quadrifoglio and Li, 2009; Zhang et al., 2017; Daganzo, 1978). Such assumptions of demand distribution are essential for the analytical models, even though they may not reflect reality well, since actual demands tend to aggregate at specific locations, e.g., train stations and shopping centers. To relax this assumption, some researchers utilized historical or actual demand to design flexible bus service (e.g., Eiró et al., 2011; Ma et al., 2017a, 2017b; Martínez et al., 2015). Amirgholy and Gonzales (2016) employed an analytical model to approximate the operating cost for running a DRT system with dynamic uniform distribution of demand. Zhang et al. (2018) developed an analytical model to investigate flexible bus service for a linear corridor connecting city center and suburb residential area. Tong et al. (2017) optimized customized bus service by considering time-space limits of the requests' origins and destinations, and solved the associated pickup and delivery vehicle routing problem with time windows (PDVRPTW) by combining various heuristics methods. R. Guo et al. (2018) solved a customized route design problem that characterized the passengers-to-vehicle assignment and compared the solutions from genetic algorithm and branch-and-cut algorithm.

As an on-demand service, the flexible bus service system should also consider the dynamics and stochasticity of demand, rather than only deterministic and historical demand. Santos and Xavier (2015) solved the dynamic DARP by matching the taxi and ride sharing service with dynamic passenger demand, while ensuring that the constraints on vehicle capacity, maximum trip cost of each passenger, and maximum delay are satisfied. Ho and Haugland (2011) endogenously considered stochastic demand and formulated the associated DARP as a two-stage stochastic integer linear program with recourse, in which a priori set of vehicle routes was determined in stage 1 to accommodate a certain set of service requests, and in stage 2, the vehicle routes determined in stage 1 were modified by skipping unrealized requests. Hyytiä et al. (2010) devised a dial-a-ride model with stochastic demand arrival and conducted analytical analyses. Q. W. Guo et al. (2018) studied stochastic dynamic switching between fixed and flexible transit services as the demand density evolves continuously over time.

The above studies, however, only considered stochasticity of demand volume and neglected the fact that the service requests are spatially stochastic and travel time varies. As flexible bus services detour to serve each service request door-to-door, the spatial stochasticity of demand and variability of travel time should not be neglected. The spatial stochasticity of demand is dependent on the clustering strategies, i.e., how to group neighborhood service requests into different zones and design the zonal flexible bus routes. In fact, a number of mathematical models were proposed to optimize flexible transit service by implementing a zonal strategy (e.g. Aldaihani et al., 2004; Li and Quadrifoglio, 2009; Pan et al., 2015; Quadrifoglio et al., 2006). Chang and Schonfeld (1991) developed optimization models for zonal service operation, which included service zone area and bus capacity as decision variables. Karabuk (2009) introduced a cluster-first route-second zonal heuristic that clusters customers to receive service together and arranges vehicle routes based on the zonal clustering. Recently, there were some studies of DRT that divided the service zone before planning, such as Lu et al. (2017) and Wang et al. (2018). In the former paper, each zone is served by one operator, and vehicles cannot traverse boundaries unless they pick up or drop off customers. In the latter research, a zone is served by only one bus, and passengers are picked up at a common terminal. By this decentralization approach, the problems were divided into several smaller sub-problems, yielding higher efficiency. However, to our best knowledge, none of these studies on DRT have jointly addressed the two aspects of demand stochasticity, namely, volume and spatial stochastic variations.

This paper aims to develop a zonal-based flexible bus system by considering both volume and spatial stochasticities of demand. The spatial stochasticity is represented by the assumption that the zonal detour times for picking up and dropping off passengers in each zone follow certain probability distributions that can be inferred from historical service requests. The detour time probability distribution inherits the variation of travel time throughout the day. Similarly, the numbers of service requests also follow certain probability distributions to reflect the volume stochasticity. The frequency-based zonal flexible bus is scheduled to minimize the operating cost, including the total regular service cost and ad hoc service cost, by considering the vehicle capacity, detour constraints within each zone, and spatial and volume stochastic variations. Naturally, it can be formulated as a two-stage stochastic problem with recourse, in which stage 1 determines the zonal visit sequences of the bus fleet and stage 2 assigns buses or ad hoc services to serve the realized service requests.

To solve this intrinsically intractable problem, we apply a service reliability (SR)-based gradient approach, which separates the original problem into two phases for solution efficiency and has been adopted for designing ferry service (An and Lo, 2014) and transit network (An and Lo, 2016), managing traffic signals with stochastic demand (Li et al., 2018), and controlling network signal with equilibrium constraint (Huang et al., 2018). Reliabilities on both volume and spatial stochasticity are introduced and optimized. The volume reliability is the probability that all service requests are picked up by the bus fleet, while the detour reliability defines the probability of the time required for a vehicle to pick up passengers.

Basically, this approach consists of two phases. In phase-1, given a specific service reliability, the number of requests to be served and the respective detour times are determined and fixed based on the historical probability distributions. The problem essentially becomes deterministic. We will then determine the fleet and vehicles' sequences of zonal visits to minimize the operating cost so as to serve the service requests. In phase-2, with the fleet and vehicles' sequences of zonal visits determined in phase-1, service requests will be assigned to vehicles subject to the detour time and vehicle capacity constraints. If certain requests are not accommodated by the regular fleet due to vehicle capacity or detour time constraints, those requests will be served by ad hoc service, which can also be considered as the cost of lost or unserved demand. The sum of phase-1 operating cost and expected phase-2 ad hoc service cost will be minimized via varying the SR by the gradient based approach. The feasible region of the two-stage stochastic problem with recourse is proven to be equivalent to the feasible region of the proposed SR-based gradient approach. In other words, any optimal solution in the original two-stage stochastic problem with recourse can be found by searching the feasible region of the SR-based formulation. This system optimizes the operating cost for operators, and the abovementioned flexible bus system is suitable for operators that provide flexible bus and taxi service, such as Didi, Uber or Grab.

The contributions of this research include (1) the concept of zonal network with intra-zonal detour is introduced and the sequences of zonal visits are optimized; (2) the formulation incorporates heterogeneous service requests, and considers the volume and spatial stochastic variations of the service requests, as well as the variation of travel time endogenously; (3) the service reliability (SR)-based gradient approach is introduced and applied to optimize the sequences of zonal visits; (4) the numerical results show that medium-size vehicles are preferred to minimize the cost while preserving the quality of service in detour time.; and (5) the formulation and solution algorithm are applied to Chengdu, China, based on real data to illustrate their applicablility.

The remainder of this paper is organized as follows. In Section 2, the two-stage stochastic formulation of the zonal-based flexible bus service is presented, and the SR-based gradient solution approach is introduced. After that, numerical examples are conducted, and policy implications are presented in Section 3. Section 4 draws the conclusion and provides future research directions.

## 2. Model Formulation

The following notation is utilized and summarized for ease of reference.

*Sets*
$Z$ — Set of zones
$V$ — Set of vehicles
$Z_v$ — Set of zones traversed by vehicle $v$
$Z_v^{RS}$ — Set of zones from zone R to zone S traverses by vehicle $v$
$P$ — Set of candidate routes
$E$ — Set of demand categories
$K$ — Set of demand scenarios
$\Omega$ — Set of OD pairs to be served
$D / D^\kappa$ — Set of service requests / Set of service requests in scenario $\kappa$

*Parameters*

| | |
|---|---|
| $\bar{t}_z$ | Maximum allowable detour time in zone $z$, $z \in Z$ |
| $\bar{t}$ | Maximum allowable detour time for the whole trip |
| $\bar{t}_{RS}$ | Maximum allowable detour time from zone R to S, $(R,S) \in \Omega$ |
| $\mathbf{T}_z^\kappa$ | A matrix of detour time for all demand in $D^\kappa$ of the vehicle in zone $z$ that considers correlations, $z \in Z, \kappa \in K$ |
| $cap$ | Vehicle capacity |
| $c_p^1$ | Operating cost of route $p$, $p \in P$ |
| $m_p$ | The number of zones traversed by route $p$ |
| $\sigma_{ijp}$ | Operating cost to provide space for a passenger from zone $i$ to zone $j$ by vehicle of route $p$, $(i,j) \in \Omega, p \in P$ |
| $c_d^2$ | The ad hoc service cost of serving request $d$, $d \in D$ |
| $p_\kappa$ | Probability of scenario $\kappa$, $\kappa \in K$ |
| $\tau_d^z$ | Detour time required in zone $z$ to serve service request $d$, $d \in D$, $z \in Z$ |
| $\tau_{db}^{z-}$ | Half of detour time reduction in zone $z$ if service request $d$ and $b$ are served simultaneously, $z \in Z, d, b \in D$ |
| $\alpha_{d+}^z$ | Binary parameter indicating service request $d$ starts in zone $z$, $d \in D$, $z \in Z$ |
| $\alpha_{d-}^z$ | Binary parameter indicating service request $d$ ends in zone $z$, $d \in D$, $z \in Z$ |
| $n_d$ | Number of passengers of service request $d$, $d \in D$ |
| $(z_{p1},...,z_{pn})$ | Zonal visit sequence for route $p$, $p \in P, z_{p1},...,z_{pn} \in Z$ |
| $\mathbf{B}_p = \{b_z^{RS}\}_p$ | A converting matrix to express the number of in-vehicle passengers when leaving zone $z$ for route $p$, where $rs$ are possible OD pairs, $(R,S) \in \Omega, z \in Z, p \in P$ |
| $\Delta_e$ | Random variable of number of service requests of demand category $e$, $e \in E$ |
| $\Lambda_z$ | Random variable of additional detour time to serve one more service request in zone $z, z \in Z$ |
| $\tilde{\tau}_z$ | Detour time required from the boundary to the first point or the last point in zone $z$ |
| $\tau_z^{II}$ | Detour time per service request served in zone $z, z \in Z$ |
| $n_e$ | Number of passengers of demand category $e$, $e \in E$ |
| $\alpha_{e+}^z$ | Binary parameter indicating demand category $e$ starts in zone $z$, $e \in E$, $z \in Z$ |
| $\alpha_{e-}^z$ | Binary parameter indicating demand category $e$ ends in zone $z$, $e \in E$, $z \in Z$ |
| $M_1, M_2$ | Very large positive numbers |

*Decision variables*

| | |
|---|---|
| $\mathbf{X} = \{x_{pv}\}$ | Binary variables indicating whether vehicle $v$ is assigned to route $p$, $p \in P, v \in V$ |
| $\mathbf{X'} = \{\chi_p\}$ | Number of vehicles assigned to route $p$, $p \in P$ |
| $\mathbf{W} = \{w_{dv}\}$ | Binary variables indicating whether service request $d$ is assigned to vehicle $v$, $d \in D, v \in V$ |
| $\mathbf{w}_v^\kappa$ | The vector of $w_{dv}$ for vehicle $v$, for all service request $d \in D^\kappa, \forall v \in V, \forall \kappa \in K$ |
| $\zeta_v^\kappa = \{\zeta_v^{RS,\kappa}\}$ | Number of passengers of OD pair RS picked up by vehicle $v$ in scenario $\kappa$, $\zeta_v^\kappa = \left[\{\zeta_v^{RS,\kappa} : (R,S) \in \Omega\}\right]^\top$, $(R,S) \in \Omega, v \in V, \kappa \in K$ |
| $\zeta_p^{RS}$ | Number of passengers from OD pair RS picked up by vehicle of route $p$, $(R,S) \in \Omega, p \in P$ |
| $\mathbf{Y} = \{y_{ev}\}$ | Number of service request of type $e$ served by vehicle $v$, $e \in E, v \in V$ |
| $\mathbf{Y} = \{y_{ep}\}$ | Number of service request of type $e$ served by vehicle with route $p$, $e \in E, p \in P$ |
| $\tilde{y}_{vz}$ | Number of pickup or drop off point of vehicle $v$ in zone $z$, $v \in V, z \in Z$ |
| $\boldsymbol{\rho}^I = \{\rho_e^I\}$ | Volume reliability, probability that all requests of service requests of type $e$ are picked up by the fleets, $e \in E$ |
| $\boldsymbol{\rho}_+^{II} = \{\rho_z^{II}\}$ | Detour reliability for service request in zone $z$. The $\rho_z^{II}$-th percentile of detour time is considered in phase-1 problem, $z \in Z$ |
| $\delta_e$ | Number of service requests of type $e$ to be picked up in phase-1, $e \in E$ |
| $C_f$ | The fixed operating cost determined in stage 1 and phase-1 |
| $\bar{Q}$ | The expected ad hoc service cost |
| $C_{total}$ | The total cost, including of operating cost and ad hoc cost |

## 2.1 Zonal-based flexible bus service

### 2.1.1 Definitions and characteristics of zone, service request and route

The service area is divided into zones geographically, as illustrated in Figure 1, which is a zonal-based network constructed with zone centroids, represented as nodes and roads connecting zone centroids as links. The zone shapes and sizes are considered as given. Indeed, the division of zones, which will affect the planning of the flexible bus service, can be optimized based on the historical demand patterns, and is our current research. The intra-zone service requests are omitted, as ZBFBS is intended for service requests whose origins and destinations fall into different zones.

Service requests are characterized by origin-destination (OD) zones, detour times, and the number of passengers per request. The calculation of detour time, considering the correlation between service requests, is explained in Section 2.3. There are two definitions of routes; first, route $p$ in this paper refers to a *zonal route*, defined as a sequence of zonal visits, and belongs to the candidate route set $P$, i.e., $p \in P$. Each zonal route has a cost $c_p^1$ that is dependent on the length of the route. The ZBFBS problem is generic for any set of zonal routes. Second, an *actual route*, which specifies the route a flexible bus actually travels including the detours measured in Euclidean distance for simplicity, which can be converted back to street network distance via certain multipliers (Yang et al., 2018). Similar to door-to-door services around the globe, drivers are flexible in determining the actual route based on the traffic condition while visiting the service requests assigned. We minimize the operator's cost of the ZBFBS by determining the route of every vehicle and assigning service requests to vehicles while ensuring the vehicle capacity constraint and detour constraint for each vehicle are satisfied.

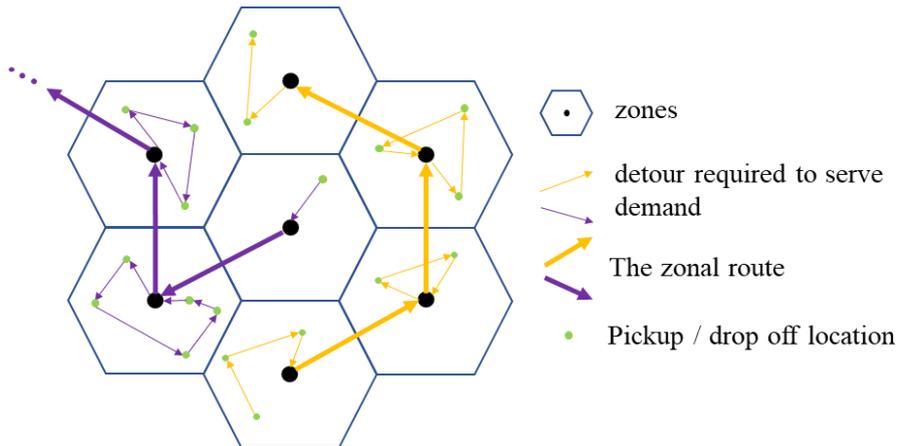

Figure 1. An illustration of zonal-based flexible bus service system

### 2.1.2 Assumptions

The following assumptions are made in this paper:

A1. The marginal detour time decreases with the number of service requests served in the zone.
A2. Door-to-door services are provided without transfers.
A3. One vehicle type is used for the regular flexible bus service.
A4. Each OD pair has at least one route traversing it.
A5. The intra-zone service requests are omitted.

## 2.2 Two-stage stochastic formulation

We aim to design a flexible zonal-based bus system while addressing the uncertainties of demand over space and volume, which follow certain probability distributions. Two types of services are provided: regular service operated with designated frequency, and ad hoc service to accommodate the demand unserved by the regular service. The stochastic formulation is divided into two stages. The first stage decides the regular service provided, where vehicles are either assigned to different routes or idle. Based on the routes determined in the first stage, the second stage matches the realized service requests with the regular service vehicles. Service requests must be served by either regular service or ad hoc service, in which the latter incurs an additional cost to the system. Note the ad hoc service provides one smaller vehicle per service request, with a predefined cost per service request. Regardless of the number of passengers to be carried per request, the entire group of passengers of each request must be served by the same vehicle.

The two-stage stochastic problem with recourse **(P0)** can be stated as,

$$\textbf{Stage-1}: \min_{\mathbf{X},\mathbf{W}} \sum_{v \in V} \sum_{p \in P} c_p^1 x_{pv} + \bar{Q}(\mathbf{X}) \tag{1}$$

subject to

$$\sum_{p \in P} x_{pv} \leq 1, \forall v \in V \tag{2}$$

$$x_{pv} \in \{0,1\}, \forall p \in P, \forall v \in V \tag{3}$$

where,

$$\textbf{Stage-2}: \bar{Q}(\mathbf{X}) = \min_{\mathbf{W}} \sum_{\kappa \in K} p_\kappa [\sum_{d \in D^\kappa} c_d^2 (1 - \sum_{v \in V} w_{dv})] \qquad (4)$$

subject to

$$\mathbf{w}_v^{\kappa \mathrm{T}} \mathbf{T}_z^\kappa \mathbf{w}_v^\kappa \leq \bar{t}_z, \forall z \in Z, \forall v \in V, \forall \kappa \in K \qquad (5)$$

$$\zeta_v^{\mathrm{RS},\kappa} = \sum_{d \in D^\kappa} n_d \alpha_{d+}^{\mathrm{R}} \alpha_{d-}^{\mathrm{S}} w_{dv}, \forall (\mathrm{R},\mathrm{S}) \in \Omega, \forall v \in V, \forall \kappa \in K \qquad (6)$$

$$x_{pv} \mathbf{B}_p \zeta_v^\kappa \leq cap \mathbf{1}_{m_p - 1}, \forall v \in V, \forall p \in P, \forall \kappa \in K \qquad (7)$$

$$\zeta_v^{\mathrm{RS},\kappa} \leq M_1 \sum_{p \in P} x_{pv}, \forall (\mathrm{R},\mathrm{S}) \in \Omega, \forall v \in V, \forall \kappa \in K \qquad (8)$$

$$\sum_{v \in V} w_{dv} \leq 1, \forall d \in D^\kappa, \forall \kappa \in K \qquad (9)$$

$$w_{dv} \in \{0,1\}, \forall d \in D^\kappa, \forall v \in V, \forall \kappa \in K \qquad (10)$$

The goal of this problem is to minimize the expected total system cost through scheduling the regular flexible bus services and introducing ad hoc services if needed in order to accommodate the realized demand. Stage 1 determines the regular zonal-based flexible bus routing. The objective function (1) minimizes the operating cost of the regular flexible bus service and expected Stage 2 cost, where $c_p^1$ is the operating cost of route $p \in P$ for a vehicle and $P$ is the set of candidate routes. $x_{pv}$ is the binary decision variable which equals 1 if vehicle $v \in V$ is assigned to route $p$, otherwise 0, forming the binary matrix $\mathbf{X} = \{x_{pv}\}$. $V$ is the set of vehicles. Vehicle can only operate on at most one route, as represented in constraint (2).

Stage 2 minimizes the expectation of the ad hoc services cost by allocating the realized demand to regular flexible buses with route determined by $\mathbf{X}$ in stage 1. Following Lo et al. (2013), we directly formulate the problem in the context of scenario simulation. Each scenario $\kappa \in K$ has a probability of $p_\kappa$ with $\sum_{\kappa \in K} p_\kappa = 1$, and a service request set $D^\kappa$. $\mathbf{W} = \{w_{dv}\}$ is the binary decision variable which equals 1 if service request $d$ is served by vehicle $v$; 0 otherwise. If a service request $d$ is not served, i.e., $\sum_{v \in V} w_{dv} = 0$, an ad hoc service cost $c_d^2$ will be incurred, as denoted in objective function (4) in Stage 2. $\mathbf{w}_v^\kappa$ is the vector form of $w_{dv}$ for all $d \in D^\kappa$. Constraint (5) enforces that the total detour time of each vehicle in any zone $z$ should not exceed the time limit $\bar{t}_z$. $\mathbf{w}_v^{\kappa \mathrm{T}}$ is the transpose of vector $\mathbf{w}_v^\kappa$. $\mathbf{T}_z^\kappa$ is a $|D^\kappa| \times |D^\kappa|$ square matrix to process the detour time considering the correlation between the service requests, a detailed explanation is in the next subsection.

Equation (6) calculates $\zeta_v^{\mathrm{RS},\kappa}$, the number of passengers from OD pair $(\mathrm{R},\mathrm{S}) \in \Omega$ served by vehicle $v$ in scenario $\kappa$, where $n_d$ is the number of passengers of request $d$. $\alpha_{d+}^{\mathrm{R}}$ and $\alpha_{d-}^{\mathrm{S}}$ indicate request $d$ starts from zone R and ends in zone S, respectively. $\Omega$ is the set of OD pairs to be served. Constraint (7) denotes the vehicle capacity constraint, which ensures that each vehicle $v$ will not exceed its maximum capacity $cap$ over the entire trip. To ensure that vehicles are within the capacity limit in the zone, vehicles first drop off some in-vehicle passengers before picking up new passengers. Vehicle capacity constraints are set for all the zones they traverse except for the end zones where no new passengers are picked up, i.e., vehicle $v$ should have $m_p - 1$ zonal capacity constraints, where $m_p$ is the number of zones in route $p$. $\zeta_v^\kappa$ is the vector of $\zeta_v^{RS,\kappa}$, and $\mathbf{B}_p = \{b_z^{RS}\}_p$ is a $(m_p - 1) \times |\Omega|$ matrix converting $\zeta_v^{RS,\kappa}$ to the number of in-vehicle passengers when vehicle $v$ leaves each zone, where $|\Omega|$ is the number of OD pairs. $\mathbf{1}_{m_p-1}$ is the column vector with $m_p - 1$ items equal to 1, indicating the $m_p - 1$ zonal capacity constraints. To construct the matrix $\mathbf{B}_p$, we first rule out all the OD pairs that are not on route $p$ and set all the corresponding $\{b_z^{RS}\}_p$ equal to a large number $M_1$, ensuring that the service requests from OD pair RS cannot be served by any vehicle on route $p$ not traversing OD pair RS. If zone $z$ is on the path from R to S, then $\{b_z^{RS}\}_p$ is equal to 1, meaning that any passengers from R to S are onboard when vehicle $v$ leaves zone $z$, and 0 otherwise. This is inspired by the capacity constraints in the classical pickup and delivery problems (Savelsbergh and Sol, 1995). Constraint (8) ensures every vehicle serving passengers has a route. Constraint (9) ensures a request to be served by at most one flexible bus. Constraint (10) defines $w_{dv} \in \mathbf{W}$ to be binary, i.e., the requests have to be served entirely regardless of the number of passengers. An illustrative example of the detour constraint, converting matrix, and vehicle capacity constraint is in Appendix A.

Note that although the detour time constraint (5) restricts the total detour time per zone, it can be replaced or supplemented by the following constraints that restrict the detour time per trip and/or per OD pair, respectively.

$$\sum_{z \in Z_v} \mathbf{w}_v^{\kappa T} \mathbf{T}_z^\kappa \mathbf{w}_v^\kappa \leq \bar{t}, \forall v \in V, \forall \kappa \in K \tag{11}$$

$$\sum_{z \in Z_v^{RS}} \mathbf{w}_v^{\kappa T} \mathbf{T}_z^\kappa \mathbf{w}_v^\kappa \leq \bar{t}_{RS}, \forall (R,S) \in \Omega, \forall v \in V, \forall \kappa \in K \tag{12}$$

$Z_v$ represents all the zones traversed by vehicle $v$, and $Z_v^{RS}$ represents the pair from zone R to zone S traversed by vehicle $v$. After determining $\mathbf{X}$ in stage 1, $Z_v$ and $Z_v^{RS}$ can be determined from the vehicle route.

## 2.3 Detour time quantification

Detour is quantified by the time required to travel between the corresponding zone centroid and the pickup or drop off location, which implicitly captures the traffic conditions. After the demand is realized, the flexible bus does not need to go back to the zone centroid after each pick-up or drop-off, but goes from one service request location to another, hence the detour time between service requests depends on the proximity between their service locations. In general, the total detour time within a zone is shorter than the sum of individual detour times of serving multiple requests (as measured from the individual request location to the zone centroid). This reduction in the total detour time within a zone is determined from the proximity between the pickup or drop-off locations.

To facilitate the calculation of detour time, after the realization of service requests, $|D^\kappa| \times |D^\kappa|$ symmetric detour time matrices $\mathbf{T}_z^\kappa$ are introduced for every zone $z$. Specifically, the $d^{\text{th}}$ element in the diagonal of $\mathbf{T}_z^\kappa$ is the detour time required in zone $z$ to pick-up or drop-off request $d$ (as measured from its location to the zone centroid), denoted by $\tau_d^z$, while the other elements are the reduction of detour time when two nearby service requests are served by same vehicle (without having to travel back to the zone centroid for each request), denoted by non-positive $\tau_{db}^{z-}$ at entry $(d,b)$. Therefore, $\tau_{db}^{z-} = \tau_{bd}^{z-}$. To ensure that the total detour time within a zone always increases when an additional service request is served, the sum of potential reductions of detour time for service request $d$ in zone $z$ must be smaller than the detour time of $d$ at $z$, formally expressed as,

$$\tau_d^z \geq \sum_{\tau_{db}^{z-} \in \mathrm{T}_d^z} -2\tau_{db}^{z-}, \forall d \in D^\kappa, \forall z \in Z, \forall \kappa \in \mathrm{K} \tag{13}$$

where $cap$ is the vehicle capacity, and $\mathrm{T}_d^z$ is a set of the $cap$-smallest $\tau_{db}^{z-}$ of $d$ for all $b \in D^\kappa$, includes the smallest and the next $(cap-1)$-smallest values of $\tau_{db}^{z-}$. If there are less than $cap$ service requests with either origin or destination in zone $z$, all the negative $\tau_{db}^{z-}$ are included in the set $\mathrm{T}_d^z$.

After the service requests are realized, $\tau_d^z$ can be obtained from the historical travel time between zone centroid and location of the service request while $\tau_{db}^{z-}$ can be calculated based on the proximity of the locations between the service requests. Note that the detour time is often overestimated as the condition (13) limits the reduction of the detour time that may be actually higher, especially when there are a lot of service requests in close proximity within a zone. An example of the detour time matrix is given in Appendix A. Calculations of detour time and detour time reduction from real data are demonstrated in Appendix G.

## 2.4 Service reliability-based formulation and solution algorithm

**P0** is a stochastic binary program that is hard to solve, especially for large numbers of demand requests or a large vehicle fleet. For the sake of computational efficiency, we decompose the problem by introducing service reliabilities, similar to An and Lo (2014) and Lo et al. (2013). The random demand volume is separated into two parts: the first part is based on the volume reliability $\boldsymbol{\rho}^{\mathrm{I}}$, which will be served by regular services; the other part is the demand above that level, which will be served by ad hoc services if needed. Similarly, the random detour time is separated by the detour time reliability $\boldsymbol{\rho}^{\mathrm{II}}$, which is considered when planning the regular services. Upon realization of the actual demand, if the total vehicle detour time exceeds the maximum allowable detour time, some service requests will be served by ad hoc services.

The volume and detour time reliabilities separate the original problem **P0** into two sub-problems **P1** and **P2**, corresponding to the phase-1 problem formulated under specified reliability measures, and the phase-2 problem formulated to assign requests to vehicles upon realization of the actual demands. **P1** tackles the aggregate service requests, defined by demand categories, while **P2** deals with individual service requests. In phase-1, we determine the schedule of regular zonal flexible bus services to cover the stochastic demand up to the specified volume reliability and detour time reliability. The number of passengers onboard should not exceed the vehicle capacity *cap* in any zone, and the total detour time per vehicle should not exceed the maximum allowable detour time $\bar{t}_1$ designated for every zone I. In phase-2, we determine the allocation of the realized service requests to the existing flexible bus determined in phase-1 and the deployment of ad hoc services. In this problem setting, the formulation is more complex than the previous studies of An and Lo (2014) and Lo et al. (2013), as it not only considers two types of stochastic variations in phase-1, namely volume and spatial stochasticities, but it also matches the service requests with specific vehicles in phase-2, while the matching process was not considered in An and Lo (2014) and Lo et al. (2013). The key to solve the problem is to determine the optimal values of the two demand reliabilities, i.e., volume reliability $\boldsymbol{\rho}^I$ representing volume stochasticity and detour time reliability $\boldsymbol{\rho}^{II}$ representing spatial stochasticity, such that the expected total operating cost is minimized.

In addition, demand categories $e \in E$ are introduced to represent the aggregate service requests $d$ by OD pairs and passenger groupings. A demand category $e$ is specified by its origin and destination, number of passengers per service request $n_e$ and volume stochasticity given by the random variable of number of requests $\Delta_e$.

### 2.4.1 Detour time approximation in phase-1

In the phase-1 problem, the calculation of detour time is separated into two parts as illustrated by arrows with different colors in Figure 2. The first part, indicated by the green arrow, is function $\tilde{\tau}_z$ that estimates the detour time between the boundary of zone $z$ and the closest service location, which is assumed to be convex decreasing with respect to $\tilde{y}_{vz}$, i.e., the number of service requests served by vehicle $v$ in zone $z$. The function is decreasing because it is more likely to have some service locations closer to the zone boundary with more service requests, hence a shorter detour. By equation (5) of **P0**, with more service requests being served, the accumulated reduction of detour time increases. Similarly, function $\tilde{\tau}_z$ captures the reduction of detour time with the number of service requests served. The total detour time is shorter than that calculated without correlation in phase-1. The second part is the detour times between locations of service requests, given by the product of $\tau_z^{II}$ and the number of trip segments, as indicated by the orange arrows between two points. $\tau_z^{II}$ is the detour time between two points specified by the detour time reliability. Similar to the diagonal values of $\mathbf{T}_z^K$ in equation (5), this part calculates the detour time per additional service request served. Moreover, the value of $\tau_z^{II}$ accounts for the proximity of the location of service requests being served, which is captured by detour time reductions in (5). For example, service request locations that are close will result in a lower $\tau_z^{II}$ in phase-1, which is equivalent to higher detour time reduction in $\mathbf{T}_z^K$.

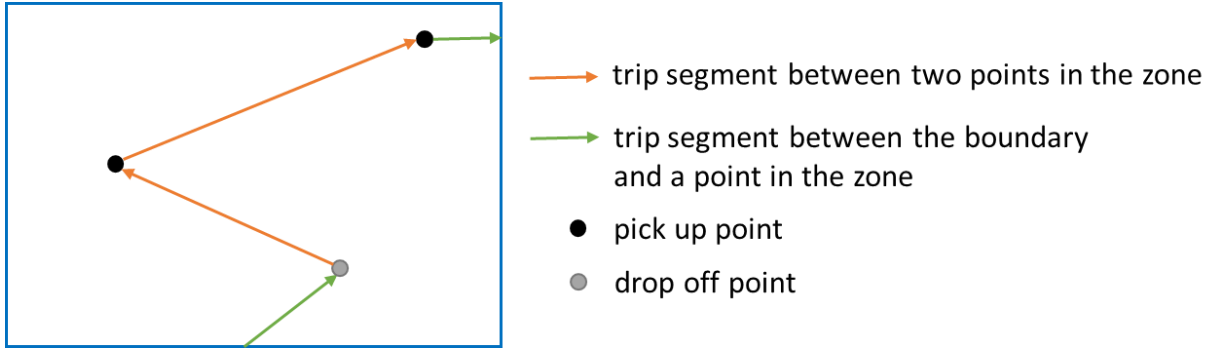

Figure 2. An example route of vehicle in a zone

By considering the detour time between the boundary and the closest service location and the detour time between the locations of two service requests, $t_{vz}$, the detour time of vehicle $v$ in zone $z$ is approximated as,

$$t_{vz} = 2\tilde{\tau}_z(\tilde{y}_{vz}) + (\tilde{y}_{vz} - 1)\tau_z^{II}, \forall v \in V, \forall z \in Z \qquad (14)$$

where $\tilde{y}_{vz}$, the number of service requests served by vehicle $v$ in zone $z$, is defined as,

$$\tilde{y}_{vz} = \sum_{e \in E}(\alpha_{e+}^z + \alpha_{e-}^z)y_{ev}, \forall v \in V, \forall z \in Z \qquad (15)$$

$y_{ev}$ is the number of service requests of demand category $e$ assigned to vehicle $v$, which is one of the decision variables in the phase-1 problem.

In the two-stage formulation, the detour times $\tau_d^z$, diagonal elements of $\mathbf{T}_z^\kappa$, capture the additional detour times for service requests, which are approximated in phase-1 by $\tau_z^{II}$ based on the detour time reliability in each zone. The detour time reductions $\tau_{db}^{z-}$ that capture the reductions of marginal detour times for the number of service requests served and proximity between the service requests are approximated in phase-1 by $\tau_z^{II}$ and $\tilde{\tau}_z$, respectively. Hence, this approximation of detour time is consistent with the two-stage formulation.

### 2.4.2 Formulation of the phase-1 problem (P1)

The phase-1 problem **P1** can be formulated as,

$$\min_{\mathbf{X},\mathbf{Y}} C_f(\boldsymbol{\rho}^I, \boldsymbol{\rho}^{II}) = \sum_{v \in V}\sum_{p \in P} c_p^1 x_{pv} \qquad (16)$$

subject to (14), (15)

$$\sum_{p \in P} x_{pv} \leq 1, \forall v \in V \qquad (17)$$

$$\delta_e = \inf\{\eta \in \mathbb{Z} \mid \Pr(\Delta_e \leq \eta) \geq \rho_e^I\}, \forall e \in E \qquad (18)$$

$$\sum_{v \in V} y_{ev} = \delta_e, \forall e \in E \qquad (19)$$

$$\zeta_v^{RS} = \sum_{e \in E} n_e \alpha_{e+}^R \alpha_{e-}^S y_{ev}, \forall (R,S) \in \Omega, \forall v \in V \quad (20)$$

$$x_{pv} \mathbf{B}_p \boldsymbol{\zeta}_v \leq cap\mathbf{1}_{m_p-1}, \forall v \in V, \forall p \in P \quad (21)$$

$$\tau_z^{II} = \inf\{\eta \mid \Pr(\Lambda \leq \eta) \geq \rho_z^{II}\}, \forall z \in Z \quad (22)$$

$$t_{vz} \leq \overline{t}_z, \forall v \in V, \forall z \in Z \quad (23)$$

$$\sum_{(R,S) \in \Omega} \zeta_v^{RS} \leq M_1 \sum_{p \in P} x_{pv}, \forall v \in V \quad (24)$$

$$x_{pv} \in \{0,1\}, \forall p \in P, \forall v \in V \quad (25)$$

$$\mathbf{Y} \geq 0 \text{ and integral} \quad (26)$$

The Phase-1 problem determines the regular flexible bus service to carry the passenger demand up the levels as specified by the volume and detour time reliability measures. Specifically, $\mathbf{X} = \{x_{pv}\}$, the assignment of vehicle $v$ to route $p$, and $\mathbf{Y} = \{y_{ev}\}$, the number of requests of demand category $e$ assigned to vehicle $v$, are optimized by fixing the volume reliability $\boldsymbol{\rho}^I$ in Equation (18) and detour time reliability $\boldsymbol{\rho}^{II}$ in (22). Equation (18) calculates the number of service requests $\delta_e$ which are to be served by the regular flexible bus service according to the volume reliability measure $\rho_e^I$ and the random variable $\Delta_e$ for each demand category $e \in E$. Similar to (18), (22) defines the detour time between two points $\tau_z^{II}$ in each zone $z$ without requiring $\tau_z^{II}$ to be integers. $\Lambda_z$ is the detour time distribution between two service points in the zone. The objective function (16) minimizes the total regular service operating cost. Constraints (20), (21) and (23) are similar to (5), (6) and (7), which denote the vehicle capacity constraints and zonal detour constraints with demand category $e$ rather than service request $d$ as expressed in (5), (6) and (7). To ensure the service requests can be served, it is reasonably assumed that $2\tilde{\tau}_z(1) \leq \overline{t}_z$ as the operator should allow enough detour time per zone to pick up or drop off at least one service request. Constraints (17) and (25) are identical with constraints (2) and (3), indicating vehicle $v$ can only be assigned at most one route $p \in P$. Equation (19) enforces that all the requests generated from (18) should be served, while they can be served separately by any vehicle $v$. Constraint (24) ensures that only vehicle assigned to a route, i.e. $\sum_{p \in P} x_{pv} = 1$, can serve customers. $M_1$ is a large positive number. Constraint (26) is similar to (10) that ensures every request is served entirely.

The LHS of constraint (21) is bilinear, so (21) can be reformulated using the linearization technique (e.g., Cen et al., 2018) as follows.

$$\mathbf{B}_p \boldsymbol{\zeta}_v \leq M_2 cap\mathbf{1}_{m_p-1} + (1-M_2)x_{pv} cap\mathbf{1}_{m_p-1}, \forall p \in P, \forall v \in V \quad (27)$$

When $x_{pv} = 0$, i.e., vehicle $v$ is not assigned to route $p$, (27) will always hold. If $x_{pv} = 1$, we have $\mathbf{B}_p \boldsymbol{\zeta}_v \leq cap\mathbf{1}_{m_p-1}$, which is the designated capacity constraint. $M_2$ is a positive number much larger than $M_1$ used in $\mathbf{B}_p$.

Furthermore, as $\tilde{\tau}_z(\tilde{y}_{vz})$ is a convex decreasing function and $\tilde{y}_{vz}$ is a natural number, (14) can be linearized by equation (28) via replacing $\tilde{\tau}_z(\tilde{y}_{vz})$ by several linear constraints (Ng et al., 2020).

$$\tilde{\tau}_z \geq (\tilde{\tau}_z(i+1) - \tilde{\tau}_z(i))(\tilde{y}_{vz} - i) + \tilde{\tau}_z(i), \forall z \in Z, \forall v \in V, \forall i \in \{0, 2, 4, ..., 2 \times cap\} \quad (28)$$

Through these adjustments, **P1** becomes a mixed-integer linear programming (MILP) problem that can be solved by commercial solvers such as IBM CPLEX.

Similar to (11) and (12), the detour time constraint (23) can be replaced or supplemented by the following constraints to account for the maximum allowable detour time in a trip,

$$\sum_{z \in Z_v} t_{vz} \leq \overline{t}, \forall v \in V \quad (29)$$

$$\sum_{z \in Z_v^{RS}} t_{vz} \leq \overline{t}_{RS}, \forall (R,S) \in \Omega, \forall v \in V \quad (30)$$

### 2.4.3 Phase-2 problem (P2)

$$\min_{\mathbf{W}} Q_\kappa = \sum_{d \in D^\kappa} c_d^2 (1 - \sum_{v \in V} w_{dv}) \quad (31)$$

$$\mathbf{w}_v^{\kappa \mathsf{T}} \mathbf{T}_z^\kappa \mathbf{w}_v^\kappa \leq \overline{t}_z, \forall z \in Z, \forall v \in V \quad (32)$$

$$\zeta_v^{RS,\kappa} = \sum_{d \in D^\kappa} n_d \alpha_{d+}^R \alpha_{d-}^S w_{dv}, \forall (R,S) \in \Omega, \forall v \in V \quad (33)$$

$$x_{pv} \mathbf{B}_p \zeta_v^\kappa \leq cap \mathbf{1}_{m_p - 1}, \forall v \in V, \forall p \in P \quad (34)$$

$$\zeta_v^{RS,\kappa} \leq M_1 \sum_{p \in P} x_{pv}, \forall (R,S) \in \Omega, \forall v \in V \quad (35)$$

$$\sum_{v \in V} w_{dv} \leq 1, \forall d \in D^\kappa \quad (36)$$

$$w_{dv} \in \{0,1\}, \forall d \in D^\kappa, \forall v \in V \quad (37)$$

The phase-2 problem is close to the Stage-2 problem in **P0**, except that the regular flexible bus deployment $\mathbf{X} = \{x_{pv}\}$ is already fixed after solving the phase-1 problem; hence constraint (34) is now linear in $\zeta_v$. The service requests set $D^\kappa$ in phase-2 is then generated from the random variable of the number of requests $\Delta_e$ across every demand category $e \in E$. Equations (32) to (37) are identical to equations (5) to (10) with only one scenario. (11) and (12) can also be used here if appropriate. Note that the calculation of detour time explained in Section 2.3 also applies here, as the locations of service requests are known in this stage. Based on the vehicle route assignment $\mathbf{X}$ solved in **P1**, the set of realized service requests $D^\kappa$ is to be assigned to vehicles, or an ad hoc service cost is incurred. It is a special case of the Stage-2 in **P0** with one scenario.

**P2** is solved for each scenario $\kappa$ generated, denote $\boldsymbol{\rho} = (\boldsymbol{\rho}^{\mathrm{I}}, \boldsymbol{\rho}^{\mathrm{II}})$ the expected ad hoc cost is given by the following function of $\boldsymbol{\rho}$,

$$\bar{Q}(\boldsymbol{\rho}) = \sum_{\kappa \in \mathrm{K}} p_\kappa Q_\kappa \qquad (38)$$

Finally, the total cost is the sum of fixed cost and ad hoc cost,

$$C_{total}(\boldsymbol{\rho}) = C_f(\boldsymbol{\rho}) + \bar{Q}(\boldsymbol{\rho}) \qquad (39)$$

With fixed volume reliability $\boldsymbol{\rho}^{\mathrm{I}}$ and detour time reliability $\boldsymbol{\rho}^{\mathrm{II}}$, $C_f$ denotes the regular flexible bus operating cost, which is the objective value in **P1**, while $\bar{Q}$ is the expected ad hoc cost solved in **P2**. The objective is to find the optimal reliability $(\boldsymbol{\rho}^{\mathrm{I}}, \boldsymbol{\rho}^{\mathrm{II}})$ that minimizes the total cost. The basic idea to solve the problem, as will be discussed later, is to develop an SR-based gradient procedure over $(\boldsymbol{\rho}^{\mathrm{I}}, \boldsymbol{\rho}^{\mathrm{II}})$ to solve **P1** and then repeatedly solve **P2** until a certain stopping criterion is reached.

### 2.4.4 Interchangeability between the two-stage and SR-based formulations

To ensure the equivalence between the two-stage and SR-based stochastic formulations, the following additional assumptions are made. It is not necessary to have these assumptions to solve the problem, but the interchangeability between these two formulations is not guaranteed without them.

B1. The possible route set $P$ contains one and only one route from each OD pair in $\Omega$, which is taken to be the shortest path.
B2. The route operating cost is additive, i.e., $c_{\mathrm{ABC}}^1 = c_{\mathrm{AB}}^1 + c_{\mathrm{BC}}^1$ (cost from zone A to zone B to zone C is equivalent to the sum of cost from zone A to zone B and zone B to zone C).
B3. Each OD pair in $\Omega$ has at least one one-person service request.

The SR-based formulation **P1-P2** is introduced by cutting the two-stage stochastic problem with recourse **P0** by the reliability measures $\boldsymbol{\rho}$. The interchangeability of **P0** and the SR-based formulations is shown by the fact that any feasible solution of **P1-P2** is feasible in **P0**; and any feasible solution of **P0** is feasible in **P1-P2** by adjusting the reliabilities. Propositions 1 and 2 below prove the result for the case of one demand category per OD, and Proposition 3 further generalizes the interchangeability to the case with multiple demand categories per OD.

**Proposition 1.** *Any feasible solution of **P1-P2** is also feasible to the original two-stage stochastic formulation **P0**.*

**Proof.** See Appendix B. □

**Proposition 2**. *For problems with the stochastic volume described by a probability distribution for any integer greater than 0, and the stochastic detour time described by a probability distribution over the range $[0, +\infty)$, if there exists only one demand category with one passenger per OD, any optimal solution of the original 2-stage stochastic formulation can be generated by the SR-based stochastic formulation.*

**Proof.** See Appendix C. □

**Proposition 3.** *For problems similar to those stated in Proposition 2 but with multiple demand categories per OD, including single passenger service requests, any optimal solution of the original 2-stage stochastic formulation can be generated by the SR-based stochastic formulation.*

**Proof.** See Appendix C. □

### 2.4.5 SR-based solution procedure

In the above subsections, the two-phase stochastic program **P1-P2** was introduced to solve the original problem **P0**. According to Proposition 1 and Proposition 3, the same global optimal solution to the original two-stage problem can be found by searching the entire feasible region of the SR-based formulation. Nevertheless, given that the problem is non-convex, this may not be practical except for very small problems. The proposed solution procedure here aims to find the descent direction in each step, thus decreases the total cost gradually to achieve a local optimal solution of reliability $\boldsymbol{\rho} = (\boldsymbol{\rho}^I, \boldsymbol{\rho}^{II})$. The ability to find local optimal solution, when the derivative of the objective function with respect to the reliability is zero, distinguishes our algorithm from heuristic methods in which solution optimality is not clearly defined. This SR-based gradient solution procedure searches for the descent direction through the service reliability vector $\boldsymbol{\rho}$. It does not involve cutting the feasible region which will lead to increases in the problem or constraint size from iteration to iteration, as in L-shaped or Multi-cut methods.

Firstly, for a set of given reliability measures $\boldsymbol{\rho}$, we obtain the respective optimal route assignment **X** by solving **P1**. Secondly, with **X** fixed in the phase-1 problem, scenarios are generated to solve **P2** and estimate the expected total cost of the phase-2 problem. Then, we approximate the partial derivative of the expected total cost with respect to the reliability measures by sensitivity analysis. The partial derivative is then used to determine the descent direction and update the reliability measures. The process is performed repeatedly until the stopping criteria are satisfied.

#### 2.4.5.1 Optimizing the step size for sensitivity analysis

In the sensitivity analysis, to approximate the partial derivative of the expected total cost with respect to the reliability measures, it is necessary to find reliability measures that produce a different phase-1 optimal solution $\mathbf{X}^*$. Otherwise, the expected total cost in phase-2 will not change. One simple way to estimate this partial derivative by perturbation analysis is to increase the reliability measure gradually by a small fixed value, such as 0.05, iteratively until the optimal solution in phase-1 changes (e.g. An and Lo, 2014; Li et al., 2018; Lo et al., 2013). However, doing so will require **P1** to be solved many times. To reduce the number of times that **P1** needs to be solved, we find the demand volumes and detour times that will make some constraints in **P1** active or binding, so that the optimal cost or $\mathbf{X}^*$ in phase-1 may change. In this section, we establish the necessary conditions that the optimal route assignment $\mathbf{X}^*$ and objective value in phase-1 may change.

An algorithm is proposed to find $\epsilon_e^I$, the maximum increment of the demand volume without violating the constraints, which is derived in Appendix D, and justified in Proposition 4. The equation (40) below calculates the maximum number of additional service requests of type *e* can be taken by vehicle $v$ that does not affect the phase-1 solution.

$$\epsilon_{ev}^{I} = \left\lfloor \min\left[ \frac{\min_{i \in Z_{ev}}(cap - \mu_{iv})}{n_e}, \frac{\bar{t}_R - t_{vR}}{\tau_R^{II}}, \frac{\bar{t}_S - t_{vS}}{\tau_S^{II}} \right] \right\rfloor \quad (40)$$

$\mu_{iv}$ is the number of passengers onboard the $i^{th}$ zone along route $p$ for vehicle $v$. $Z_{ev} \subseteq Z$ is the set of visited zones of vehicle $v$ (ending zone excluded) for any service request of $e$. $\tilde{y}_{vR}^{*}$ is calculated from $\mathbf{Y}^{*}$ in the previous solution by (15). The complete procedure to find $\epsilon_{e}^{I}$ is formally described in Algorithm 1.

**Algorithm 1. Finding the maximum demand volume increment**

Input: Current optimal route assignment $\mathbf{X}^{*}$, passenger assignment $\mathbf{Y}^{*}$, route set $P$, vehicle set $V$, and demand category $e \in E$.
Step 0: Initialize the maximum demand volume increment $\epsilon_{e}^{I} \leftarrow 0$.
Step 1: Find the set of all routes $P_e \subseteq P$ that is possible to serve $e$.
Step 2: **For** every route $p \in P_e$
　　　Step 2.1: **For** every vehicle $v \in V$ such that $x_{pv} = 1$
　　　　　Step 2.1.1: Calculate the bound $\epsilon_{ev}^{I}$ by equation (40)
　　　　　Step 2.1.2: $\epsilon_{e}^{I} \leftarrow \epsilon_{e}^{I} + \epsilon_{ev}^{I}$
Output: $\epsilon_{e}^{I}$

**Proposition 4.** The objective value of **P1** will not change if the volume of demand category $e$, $\delta_{e}^{I}$, is increased by any $0 \leq \Delta\delta_{e}^{I} \leq \epsilon_{e}^{I}$, where $\epsilon_{e}^{I}$ is the maximum demand volume increment given by Algorithm 1.

**Proof.** See Appendix E. □

Similarly, Proposition 5 shows that the objective function will not change if the travel time between requests is increased by any non-negative value less than $\epsilon_{z}^{II}$ for zone $z$ given passenger to vehicle assignment $\mathbf{Y}^{*} = \{y_{ev}^{*}\}$ as defined below. The superscript II is the same indicator as detour reliability $\rho^{II}$, denoting detour time. $\varepsilon$ is a small number to avoid division by zero.

$$\epsilon_{z}^{II} = \min_{v \in V} \frac{\bar{t}_z - t_{vz}}{\tilde{y}_{vz}^{*} + \varepsilon} \quad (41)$$

**Proposition 5.** Given passenger assignments to vehicle $\mathbf{Y}^{*}$, for any demand category $e$, the objective value of **P1** does not change if the travel time between requests of zone $z$ is increased by any $\Delta\tau_{z}^{II} \leq \epsilon_{z}^{II}$

**Proof.** See Appendix E. □

Algorithm 1 and equation (41) determine the step size in the perturbation analysis of partial derivative more effectively, especially when the variance of the volume or detour time probability distributions are small such that slight changes of reliability measures do not change the optimal solution in phase-1. However, although they are necessary conditions for invoking change in the objective function, they are not sufficient conditions as they only detect when the constraints will become active with fixed $\mathbf{X}^*$ and $\mathbf{Y}^*$. Moreover, even rendering these constraints active will change $\mathbf{X}^*$ and/or $\mathbf{Y}^*$, the objective value may still not change. Nevertheless, it improves the efficiency of the algorithm, especially when it takes a long time to solve **P1**. It is used in Step 5.1.1 and 5.1.4 in Algorithm 2.

### 2.4.5.2 Detailed solution procedure

Recently, various methods for stochastic gradient descent optimization have been developed, such as AdaGrad (Duchi et al., 2011), RMSProp (Hinton et al., 2012), Adam (Kingma and Ba, 2014), etc., as reviewed in Ruder (2016). In this paper, we solve the problem by a mini-batch gradient descent optimization, combining the gradient solution developed in An and Lo (2014), and the Adam method. The Adam method will be used only if the objective value does not improve in three consecutive iterations, as indicated by $A=1$ if Adam method is applied, and $A=0$ if the analytical method is used. To further enhance the computation efficiency, a decomposition approach to reduce the problem size is introduced in Appendix F.

**Algorithm 2. Service reliability based gradient descent optimization**

Step 1: Set $k=1$, initialize $\boldsymbol{\rho}_k = (\boldsymbol{\rho}_k^{\mathrm{I}}, \boldsymbol{\rho}_k^{\mathrm{II}})$, where subscript $k$ denotes the iteration indicator, and initialize a binary variable $A$ as 0, indicating that the *Adam method* is not used.

Step 2: Based on $\boldsymbol{\rho}_k$, solve **P1** to find the optimal $\mathbf{X}, \mathbf{Y}$.

Step 2.1: If **P1** has no feasible solution, the value of $\boldsymbol{\rho}_k$ is reduced until a feasible solution is found in **P1**.

Step 3: Given the vehicle routes $\mathbf{X}$, generate service request scenarios, and assign the service requests to vehicles by **P2**. Calculate the expected ad hoc cost by (38) and the total cost $C(\boldsymbol{\rho}_k)$ by (39); set the optimal objective value $C^*$ as $C(\boldsymbol{\rho}_k)$ if $C^*$ does not exist or $C^* > C(\boldsymbol{\rho}_k)$.

Step 3.1: If the percentage change of objective value is less than a threshold, say 1%, terminate.

Step 4: If the optimal objective value $C^*$ does not improve in three consecutive iterations and $A$ is 0, set $A$ as 1, and *Adam method* will be used in Step 5.2 to determine the step size.

Step 5: Optimize $\boldsymbol{\rho}$.

Step 5.1: Calculate the partial derivative of $C(\boldsymbol{\rho})$ with respect to $\boldsymbol{\rho}^{\mathrm{I}}$ and $\boldsymbol{\rho}^{\mathrm{II}}$ through the perturbation analysis. As $\mathbf{X}$ is binary, it is impossible to analytically find the partial derivatives of $C(\boldsymbol{\rho})$ with respect to $\boldsymbol{\rho}$. Therefore, we use perturbation analysis to find the partial derivative as shown in the procedure below.

Step 5.1.0: Given $\boldsymbol{\rho}_k$ and its corresponding solution $(\mathbf{X}_k, C(\boldsymbol{\rho}_k))$

Step 5.1.1:
  (a) Run Algorithm 1 to obtain the maximum demand volume increment $\epsilon_e^{\mathrm{I}}$ such that the constraints are not violated. Set $\rho_{e'k}^{\mathrm{I}}$ by inverting $\Delta_e$ such that $\delta_e \leftarrow \delta_e + \epsilon_e^{\mathrm{I}} + 1$, other reliabilities being the same as $\boldsymbol{\rho}_k$, denoted as $\boldsymbol{\rho}_{e'k}^{\mathrm{I}}$.
  (b) Solve **P1** again to obtain the route assignment $\mathbf{X}_{e'k}^{\mathrm{I}}$, if the objective value of **P1** changes, go to Step 5.1.2; otherwise, go back to Step 5.1.1(a).

Step 5.1.2: Check if $\mathbf{X}_{e'k}^{\mathrm{I}}$ is in the solution set or not and get the objective value $C$ directly if it does; otherwise solve **P2** repeatedly to obtain $C$, calculate the sensitivity of $\rho_{ek}^{I}$: $\dfrac{\Delta C(\boldsymbol{\rho}_k)}{\Delta \rho_{ek}^{I}} \leftarrow \dfrac{C(\boldsymbol{\rho}_{e'k}^{I}) - C(\boldsymbol{\rho}_k)}{\rho_{e'k}^{I} - \rho_{ek}^{I}}$. Save $(\mathbf{X}_{e'k}^{I}, C(\boldsymbol{\rho}_{e'k}^{I}))$ into the solution set for future usage.

Step 5.1.3: Repeat step 5.1.1 to 5.1.2 for every demand category $e$ in $\boldsymbol{\rho}^{I}$ to obtain sensitivities over $\boldsymbol{\rho}_k^{I}$, denoted by the vector $\nabla C(\boldsymbol{\rho}_k^{I})$.

Step 5.1.4:
    (a) Obtain the detour time increment $\epsilon_z^{II}$ such that the constraints are not violated by equation (41). Set $\rho_{z'k}^{II}$ such that $\tau_z^{II} \leftarrow \tau_z^{II} + \epsilon_z^{II} + \varepsilon$ by inverting $\Lambda_z$, other reliability measures being the same as $\boldsymbol{\rho}_k$, denote it as $\boldsymbol{\rho}_{z'k}^{II}$. $\varepsilon$ is a very small value to violate some constraint.

    (b) Solve **P1** again to obtain the route assignment $\mathbf{X}_{z'k}^{II}$, if the objective value changes, go to Step 5.1.5; otherwise, go back to Step 5.1.4(a).

Step 5.1.5: Check if $\mathbf{X}_{z'k}^{II}$ is in the solution set or not, get the objective value $C$ directly if it does; otherwise solve **P2** repeatedly to obtain $C(\boldsymbol{\rho}_{z'k}^{II})$, calculate the sensitivity of $\rho_{zk}^{II}$: $\dfrac{\Delta C(\boldsymbol{\rho}_k)}{\Delta \rho_{zk}^{II}} \leftarrow \dfrac{C(\boldsymbol{\rho}_{z'k}^{II}) - C(\boldsymbol{\rho}_k)}{\rho_{z'k}^{II} - \rho_{zk}^{II}}$. Save $(\mathbf{X}_{z'k}^{II}, C(\boldsymbol{\rho}_{z'k}^{II}))$ into the solution set for future usage.

Step 5.1.6: Repeat Step 5.1.4 to 5.1.5 for every zone $z$ in $\boldsymbol{\rho}^{II}$ to obtain sensitivities of $C(\boldsymbol{\rho}_k)$ over $\boldsymbol{\rho}_k^{II}$, denoted by the vector $\nabla C(\boldsymbol{\rho}_k^{II})$.

Step 5.1.7: Join the vector $\nabla C(\boldsymbol{\rho}_k^{I})$ and $\nabla C(\boldsymbol{\rho}_k^{II})$ to become a vector of all reliabilities $\nabla C(\boldsymbol{\rho}_k)$

Step 5.2: Determine the step size and update the reliability. If $A$ is 0, go to step 5.2.1. Otherwise, go to Step 5.2.2.

  Step 5.2.1: Update the reliability by the same way as in An and Lo (2014).

    Step 5.2.1.1: Take the negative sensitivity vector $-\nabla C(\boldsymbol{\rho}_k)$ as the descent direction. The step size $\pi_k$ is chosen as: $\pi_k \leftarrow \lambda_k \dfrac{C(\boldsymbol{\rho}_k) - \gamma C^*}{\|\nabla C(\boldsymbol{\rho}_k)\|^2}$.

    Step 5.2.1.2: Update the reliability $\boldsymbol{\rho}$ in next iteration. Go to Step 5.2.3.

  Step 5.2.2: Apply Adam method (Kingma and Ba, 2014) to optimize reliability.

    Step 5.2.2.1: If it is initialized, go to Step 5.2.2.2. Otherwise, set two parameters $\mathbf{m}_{k-1} \leftarrow \mathbf{0}$ and $\mathbf{v}_{k-1} \leftarrow \mathbf{0}$, $\epsilon$ is set to be a very small number, say $10^{-6}$.

    Step 5.2.2.2: Calculate $\mathbf{m}_k \leftarrow \beta_1 \mathbf{m}_{k-1} + (1-\beta_1)\nabla C(\boldsymbol{\rho}_k)$

    Step 5.2.2.3: Calculate $\mathbf{v}_k \leftarrow \beta_2 \mathbf{v}_{k-1} + (1-\beta_2)\nabla C(\boldsymbol{\rho}_k) \odot \nabla C(\boldsymbol{\rho}_k)$

    Step 5.2.2.4: Calculate $\hat{\mathbf{m}}_k \leftarrow \mathbf{m}_k / (1-\beta_1^k)$

    Step 5.2.2.5: Calculate $\hat{\mathbf{v}}_k \leftarrow \mathbf{v}_k / (1-\beta_2^k)$

    Step 5.2.2.6: Update the reliability $\boldsymbol{\rho}$ in next iteration by $\boldsymbol{\rho}_{k+1} = \boldsymbol{\rho}_k - \alpha_k \hat{\mathbf{m}}_k \oslash (\sqrt{\hat{\mathbf{v}}_k} + \epsilon)$, an elementwise operation of each reliability.

  Step 5.2.3: If any element in $\boldsymbol{\rho}_{k+1} < 0$ or any element in $\boldsymbol{\rho}_{k+1} \geq 1$, project that element onto [0,1), set $k = k+1$. Go to Step 2.

$\lambda_k$, $\gamma$, $\alpha_k$, $\beta_1$ and $\beta_2$ are parameters determined by trial-and-error. $\beta_1^k$ and $\beta_2^k$ are the $k^{\text{th}}$ power of $\beta_1$ and $\beta_2$, respectively. $\odot$ denotes element-wise multiplication (also called the Hadamard product), and $\oslash$ denotes the element-wise division of two vectors (also called the Hadamard division). The solution approach outlined above will stop if the stopping criteria is matched, resulting in an optimal **X** and total cost $C^*$.

## 3. Numerical examples

**P1**, **P2** and the algorithms are implemented by Python 3.7 PICOS API, and then solved by IBM CPLEX 12.10. The formulation will first be demonstrated by a small example in Section 3.1. Then, a five-zone scenario is introduced in Section 3.2, and solved by Algorithm 2 introduced in Section 2.4.5. Subsequently, sensitivity analyses on the vehicle capacity and detour limit are conducted based on the five-zone scenario in Section 3.3. The computation times under different problem sizes are analyzed in Section 3.4. Finally, a case study based on real data in Chengdu, China, is illustrated in Section 3.5.

### 3.1 Small illustrative example on stochastic volume and detour time

We illustrate the ZBFBS formulation in a small area with three zones, with their stochastic demand characteristics shown in Table 1. There is one and only one demand category in OD pair AC. Truncated normal distribution is used to describe the volume and detour time distributions, denoted as $TN(\mu, \sigma^2, a, b)$ that a normal distribution of mean $\mu$ and standard deviation $\sigma$ are bounded by $[a,b]$. As the truncated normal distribution is a continuous distribution, the number of service requests generated is rounded off to the nearest integer. Note that the proposed formulation and algorithm can handle any discrete distribution of integer for the demand distribution, and any continuous distribution for the detour time distribution. Moreover, there are 25 vehicles available, each with vehicle capacity $cap = 12$. The ad hoc cost $c_d^2$ to serve a request is 90% of the operating cost of one vehicle (i.e., if service request $d$ is from OD pair IJ, $c_d^2 = 0.9 c_{IJ}^1$). $c_{IJ}^1$ is the operating cost of the direct route from I to J for a flexible bus and $c_d^2$ is the ad hoc cost of service request $d$. The ad hoc cost per service request is much higher than the regular service operating cost per service request, as vehicles are shared in regular service, but not in ad hoc service. The maximum allowable detour time is 8 minutes per zone. The detour times from the boundaries of zone A and C are given by $\tilde{\tau}_A(\tilde{y}_{vz}) = 0.7 - 0.03 \tilde{y}_{vz}$ and $\tilde{\tau}_C(\tilde{y}_{vz}) = 0.6 - 0.03 \tilde{y}_{vz}$ respectively, $\tilde{y}_{vz}$ is the number of service requests served by vehicle $v$ in zone $z$. In phase-2, the detour time reduction for the detour matrix $\mathbf{T}_z$ is calculated by,

$$\tau_{db}^{z-} = -\min\left(\tau_d^z, \tau_b^z\right)/24 \tag{42}$$

$\tau_d^z$ and $\tau_b^z$ are the detour times required in zone $z$ to serve service request $d$ and $b$ respectively, generated from the detour time distribution $\Lambda_z$. The denominator 24 is to fulfill condition (13) as the maximum vehicle capacity in the numerical examples is 12.

Table 1. Demand categories and characteristics

| Demand category id ($e$) | OD pair | Demand volume distribution ($\Delta_e$) | Zone A detour time distribution ($\Lambda_A$) | Zone C detour time distribution ($\Lambda_C$) | Number of passengers per request ($n_e$) |
|---|---|---|---|---|---|
| 1 | AC | $TN(16,6,0,+\infty)$ | $TN(1.5,1,0,+\infty)$ | $TN(1,1,0,+\infty)$ | 1 |

The shortest route is ABC, with an operating cost of HK$10. As the problem size is small, the **P1** solutions can be worked out by enumeration. Specifically, reliability measures from 0 to 1, discretized by 0.05 intervals, are enumerated for the stage 1 problem. The expected ad hoc cost in **P2** is taken as the average of optimized ad hoc cost of 150 randomly generated scenarios based on the above demand profile.

It is observed that there are multiple optimal reliability measures leading to the same best route design, such as $\boldsymbol{\rho}=(\rho_1^I,\rho_A^{II},\rho_C^{II})=(0.3,0.25,0.3)$ or $\boldsymbol{\rho}=(0.4,0.2,0.25)$, in which the expected total cost is HK$24.1 and two vehicles are assigned to route ABC. The former solution gives a higher reliability measure to origin detour, but lower reliability measures to demand volume and destination detour, while the latter solution produces a higher reliability measure in demand volume and destination detour. The second-best deployment has higher reliability measures, e.g., $\boldsymbol{\rho}=(0.4,0.45,0.4)$ which assigns three vehicles to route ABC with the expected total cost equal to HK$30.0. This enumeration also shows the trade-off between the regular service operating cost and the ad hoc service cost. Reliability measures that are too high provide excessive regular services that will increase the fixed operating cost $C_f$, but reliability measures that are too low will increase the ad hoc cost $\bar{Q}$ to serve the excessive demand. The best solution comes from a reliability measure that balances the trade-off between operating cost and variable ad hoc cost, as illustrated in Figure 3.

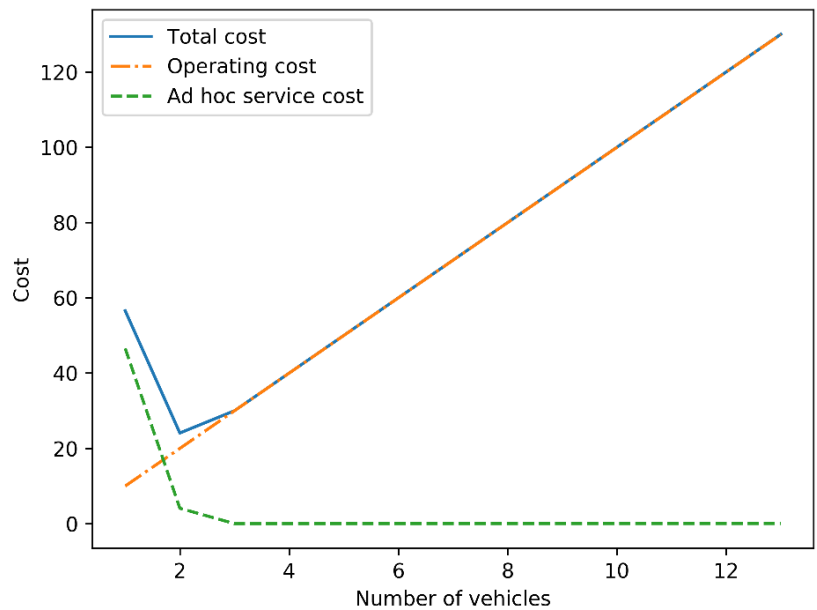

Figure 3. The costs with respect to number of vehicles in this example

The number of solutions is limited due to the integral constraints of vehicle to route assignment **X**. In other words, for some different reliability measures, the same **X** is generated. Therefore, there are multiple reliability measures that yield the same global optimal solution. Moreover, the total cost $C$ is non-convex with respect to $\boldsymbol{\rho}$. To show the non-convexity, we enumerate the volume reliability measure $\rho^I$ from 0 to 1, discretized by 0.001 intervals, and solve **P1-P2** for each volume reliability measure enumerated, given detour time reliability measures $\rho_A^{II}$ and $\rho_C^{II}$ as 0.7. The result, depicted in Figure 4, illustrates the non-convexity of the total cost with respect to volume reliability. It is shown that the total cost increases rapidly at some points where one more vehicle is required to serve the service requests. Therefore, convex optimization approaches that are well-studied in the literature are not applicable to this problem.

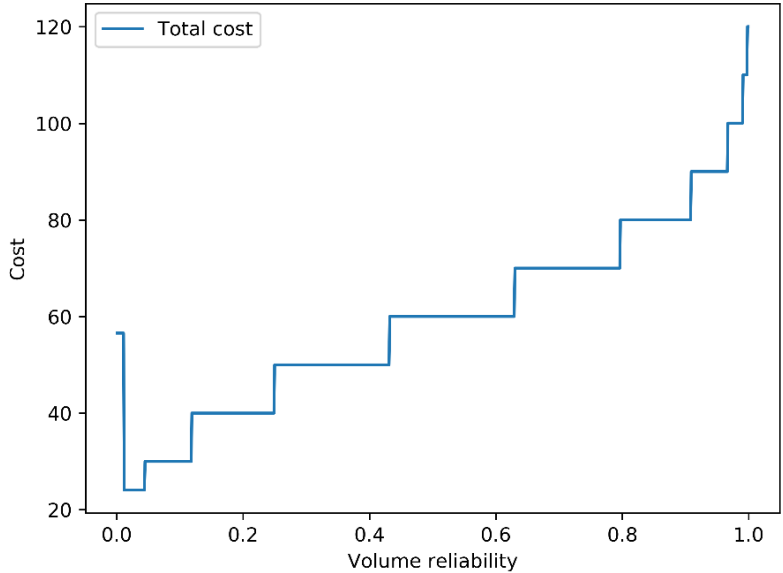

Figure 4. The score with respect to $\rho^{\mathrm{I}}$, where $\rho_{\mathrm{A}}^{\mathrm{II}} = \rho_{\mathrm{C}}^{\mathrm{II}} = 0.7$

## 3.2 Numerical examples in a five-zone scenario

To demonstrate the capability of the SR-based gradient solution approach, we divide the service area into five zones as shown in Figure 5. The blue dots denote their centroids, and the blue lines are the links connecting the zones, where the blue numbers are the link costs per 10-seater vehicles in HK$. The examples are performed on a desktop computer with Intel i7-3770 3.4Ghz CPU and 32Gb ram. The decomposition outlined in Appendix F is applied to this and subsequent examples to reduce the solving time without loss of generality.

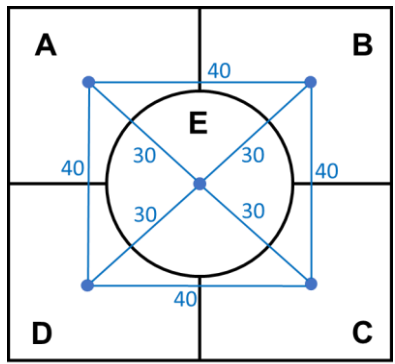

Figure 5. The five-zone network topology

The demand categories, with their demand volume distributions and numbers of passengers, are given in Table 2. The probability distributions of the travel times between two service locations are given in Table 3. These probability distributions are also used to generate diagonal values of $\mathbf{T}_z$, the matrix of detour time for each zone $z$. The correlations of the detour time, or detour time reductions, are calculated by (42) as in the first example, at the beginning of Section 3.1.

Table 2. Volume stochasticity for the five-zone network

| OD | Demand volume distribution ($\Delta_e$) | Number of passengers ($n_e$) | OD | Demand volume distribution ($\Delta_e$) | Number of passengers ($n_e$) |
|---|---|---|---|---|---|
| AB | $TN(3,4,0,+\infty)$ | 1 | DE | $TN(5,4,0,+\infty)$ | 1 |
| BA | $TN(3,4,0,+\infty)$ | 2 | ED | $TN(4,9,0,+\infty)$ | 2 |
| AC | $TN(3,4,0,+\infty)$ | 1 | CE | $TN(2,4,0,+\infty)$ | 1 |
| BC | $TN(3,4,0,+\infty)$ | 3 | EC | $TN(2,1,0,+\infty)$ | 2 |
| AD | $TN(4,4,0,+\infty)$ | 2 | EB | $TN(2,9,0,+\infty)$ | 1 |
| DA | $TN(4,4,0,+\infty)$ | 1 | DB | $TN(6,1,0,+\infty)$ | 1 |
| AE | $TN(4,9,0,+\infty)$ | 1 | BD | $TN(4,4,0,+\infty)$ | 3 |
| EA | $TN(2,1,0,+\infty)$ | 3 | CB | $TN(5,9,0,+\infty)$ | 2 |
| BE | $TN(6,16,0,+\infty)$ | 2 | EA | $TN(2,1,0,+\infty)$ | 1 |
| CD | $TN(3,4,0,+\infty)$ | 1 | | | |

Table 3. The detour time distributions used for zones

| Zone | Detour time for the segment between two points ($\Lambda_z$) | Detour time from the boundary ($\tilde{\tau}_z(\tilde{y}_{vz})$) |
|---|---|---|
| A | $TN(1.5,1,0,+\infty)$ | $0.6e^{-\tilde{y}_{vz}/12}$ |
| B | $TN(1,1,0,+\infty)$ | $0.4e^{-\tilde{y}_{vz}/12}$ |
| C | $TN(1,1,0,+\infty)$ | $0.4e^{-\tilde{y}_{vz}/12}$ |
| D | $TN(1,1,0,+\infty)$ | $0.4e^{-\tilde{y}_{vz}/12}$ |
| E | $TN(1.5,1,0,+\infty)$ | $0.6e^{-\tilde{y}_{vz}/12}$ |

Under assumption B1, only the shortest route serving each OD is considered, so the candidate route set P includes routes AB, BA, AEC, BC, AD, DA, AE, BE, CD, DE, ED, CE, EC, EB, DEB, BED, CB, and EA. The operating cost is given by summing link costs that a route traverses. A fleet of 80 10-seater vehicles and maximum detour of 8 minutes per zone per vehicle are considered. The problem is solved by the SR-based gradient solution approach outlined in Section 2.4.5. The stopping criterion is set to be 1% of the relative difference between the objective values. $\lambda_k = 0.005$, $\gamma = 0.05$, $\beta_1 = 0.9$, $\alpha_k = 0.2$ and $\beta_2 = 0.999$. The ad hoc cost is 90% of the operating cost of one vehicle, the same as the small illustrative example.

### 3.2.1 Convergence of SR-based gradient solution approach

As this stochastic formulation is inherently non-convex, different initial values of $\boldsymbol{\rho}$ to start the solution procedure may yield different solutions. To compare the effects of different initial points, we initialize the solution approach with different service reliability measures. Every entry in the initialized $\boldsymbol{\rho}$ is enumerated by roughly an interval of 0.1. The optimal solutions resulted from different initial points are plotted in Figure 6. The objective values are represented by the blue lines as the iteration proceeds. Vertical grid lines separate different initial points of service reliabilities, as indicated at the top of each box. The width between each vertical grid line represents the number of iterations, also indicated by the number at the end of each blue line. For instance, for the initial point of $\boldsymbol{\rho} = 0.6$, it takes 5 iterations to converge and the local minimum is HK$952.1. The best local minimum is HK$919.5 with the initial point $\boldsymbol{\rho} = 0.5$.

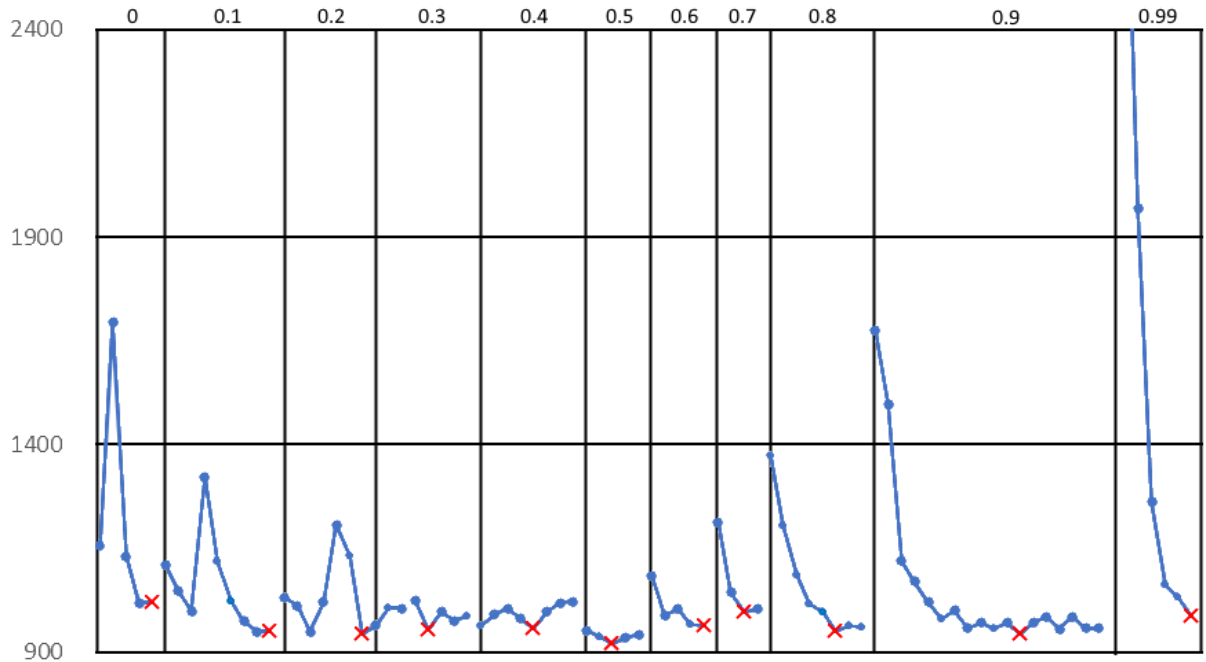

Figure 6. Objective function values versus iteration numbers for different initial points

### 3.2.2 Comparison between deterministic and reliability-based approaches

In this section, we compare the reliability-based approach with the deterministic approach to find the optimal solution by simply making use of the mean and median of the distributions as point estimates to plan for vehicle zonal routings and assignments, while ignoring the stochastic nature of demand. We then evaluate the deterministic solution by simulating its performance on 200 randomly generated demand scenarios and record its computation time and ad hoc service cost. The objective functions are evaluated and compared with the optimal solutions of the stochastic formulation with different starting points as in Section 3.2.1. Given the current volume and detour time distributions, the computing time is reduced by using deterministic approach and the resultant total cost obtained by the reliability-based approach is lower by about 5%. Nevertheless, if the distribution is not truncated normal distribution, which is sort of symmetric, but follows a lognormal distribution as follows, which is more common, the results are very different:

$$f(x) = \frac{1}{0.8x\sqrt{2\pi}} \exp\left(\frac{-\log^2(x/4)}{1.28}\right), x > 0 \qquad (43)$$

The computation times and total costs according to the deterministic and reliability-based approaches under lognormal detour time distributions are shown in Table 4.

Table 4. Computation time and cost using deterministic and reliability-based approach

|  | Mean result | Median result | Optimal reliability-based result |
|---|---|---|---|
| Computation time (s) | 581 | 1138 | 3682 |
| Number of iterations | - | - | 4 |
| Vehicle used | 40 | 40 | 18 |
| Operating cost (HK$) | 1820 | 1790 | 820 |
| Ad hoc service cost (HK$) | 970 | 952 | 1359 |
| Total cost (HK$) | 2790 | 2742 | 2179 |

The results demonstrate the capability of the reliability-based solution approach in finding good solutions for all sorts of distributions, while the deterministic approach may not. In the case of the demand following the lognormal distribution, as shown in Table 4, the reliability-based approach reduces the total cost by 20.5%, from HK$2742 to HK$2179, as compared with the deterministic approach.

### 3.3 Sensitivity analysis on vehicle capacity and maximum allowable detour time

In the planning stage, there is a trade-off between economies of scale of capacity and expected occupancy. A larger vehicle yields a lower operating cost per space due to the same fixed costs such as driver cost, vehicle registration fee, but seats are wasted if insufficient passengers are picked up. This section conducts a sensitivity analysis on the vehicle capacity and maximum allowable detour time. Four scenarios with vehicle capacity of 7, 8, 10, 12 are considered, with operating cost factored by 85%, 90%, 100% and 110%, respectively. If a ride request is not served by regular service, an ad hoc service, with a cost of 80% of the link cost shown in Figure 5, is assigned to serve that request. The maximum allowable detour time is 8 minutes. Analyses in this section are conducted by repetitively solving **P2** 200 times for the best routing **X** obtained in the SR-based gradient solution approach. The average detour times and vehicle occupancy across all scenarios are taken, as shown in Table 5 and illustrated in Figure 7.

Table 5. Results under different vehicle capacities and operating costs

|  | Scenario 1 | Scenario 2 | Scenario 3 | Scenario 4 |
|---|---|---|---|---|
| Vehicle capacity *cap* | **7** | **8** | **10** | **12** |
| Factor of operating cost | 0.85 | 0.9 | 1 | 1.1 |
| Total cost (HK$) | 980.7 | 965.7 | 939.6 | 980.8 |
| Vehicle operating cost $C_f$ (HK$) | 799 | 657 | 810 | 671 |
| Ad hoc cost (HK$) | 181.7 | 308.7 | 129.6 | 309.8 |
| Average volume reliability $\bar{\rho}^{I}$ | 0.4356 | 0.2879 | 0.4789 | 0.2008 |
| Average detour reliability $\bar{\rho}^{II}$ | 0.2331 | 0.1682 | 0.5 | 0.0339 |
| Vehicle used | 20 | 16 | 18 | 14 |
| Vehicle occupancy | 63.9% | 66.3% | 53.9% | 56.8% |
| Total detour time (mins) | 175.2 | 162.9 | 174.3 | 152.2 |
| Average maximum detour time (mins) | 4.92 | 5.40 | 5.42 | 5.77 |
| Detour time per zone visit (mins) | 3.97 | 4.42 | 4.51 | 4.95 |
| Computation time(s) | 2218 | 3308 | 1984 | 1074 |
| Number of iterations | 8 | 9 | 2 | 9 |

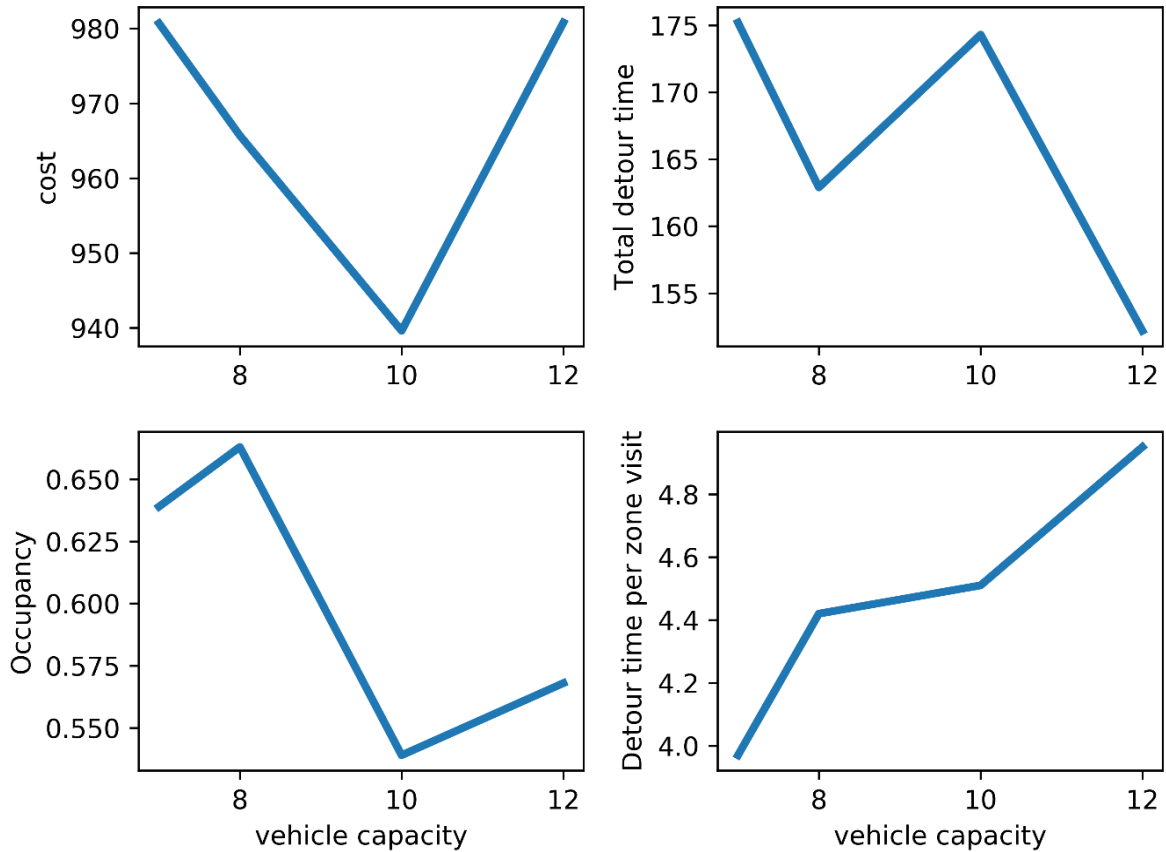

Figure 7. Total cost, vehicle occupancy and detour time versus vehicle capacity

In terms of operator cost, 10-seaters are lower than all other vehicle types. However, 7-seaters perform the best from customers' perspective, due to their lowest detour time per zonal visit (3.97 minutes) as compared with 8-seaters (4.51 minutes), 10-seaters (4.42 minutes), and 12-seaters (4.95 minutes). The vehicle occupancy is generally higher with a smaller-size vehicle. The total cost is higher for vehicle capacity of 7 and 12 and is the lowest for the vehicle capacity of 10, as it provides flexibility while saving the operating cost. In terms of occupancy, vehicles of capacity 10 and 12 yield lower occupancy than vehicles of capacity 7 and 8.

The maximum detour times for all vehicle types are far lower than the detour time constraint (set to be 8 minutes) in all three scenarios, indicating that detour time is usually not a binding constraint. The minimum slack in detour time, i.e. the maximum allowable detour time minus the average maximum detour time, is 2.23 minutes across all scenarios at Scenario 4. The maximum detour minute decreases from 5.77 to 4.92 with the usage of smaller vehicle since fewer passengers are picked up per zone on average. Note that the occupancy and total detour time are non-monotonic due to the amount of vehicles used. For example, as there are 18 flexible buses assigned in Scenario 3, i.e., more seats are provided, so that the occupancy is lower than the case when 12-seaters are used.

The discussion above poses another question: How will the service patterns change by varying the maximum allowable detour time constraint, from 6, 7, 9, to 10 minutes? The results are summarized in Table 6. Total cost, vehicle occupancy, total detour time and detour time in zone versus detour time limit are illustrated in Figure 8.

Table 6. Detour and occupancy under different detour time limits

|  | Scenario 5 | Scenario 6 | Scenario 3 | Scenario 7 | Scenario 8 |
|---|---|---|---|---|---|
| Detour time limit (mins) | **6** | **7** | **8** | **9** | **10** |
| Vehicle capacity | **10** | **10** | **10** | **10** | **10** |
| Total cost (HK$) | 1119.3 | 988.2 | 939.6 | 887.0 | 882.9 |
| Vehicle operating cost $C_f$ (HK$) | 760 | 730 | 810 | 670 | 670 |
| Ad hoc cost (HK$) | 326.3 | 258.2 | 129.6 | 217.0 | 212.9 |
| Average volume reliability $\bar{\rho}^{I}$ | 0.4887 | 0.4298 | 0.4789 | 0.4473 | 0.4183 |
| Average detour reliability $\bar{\rho}^{II}$ | 0.1985 | 0.2532 | 0.5 | 0.2930 | 0.2706 |
| Vehicle used | 17 | 16 | 18 | 15 | 15 |
| Vehicle occupancy | 51.9% | 55.0% | 53.9% | 62.0% | 63.3% |
| Total detour time (mins) | 155.4 | 162.9 | 174.3 | 161.9 | 167.4 |
| Detour time per zone visit (mins) | 4.03 | 4.51 | 4.51 | 4.88 | 5.04 |
| Computation time(s) | 10788 | 5000 | 1984 | 1320 | 733 |
| Number of iterations | 4 | 3 | 2 | 4 | 3 |

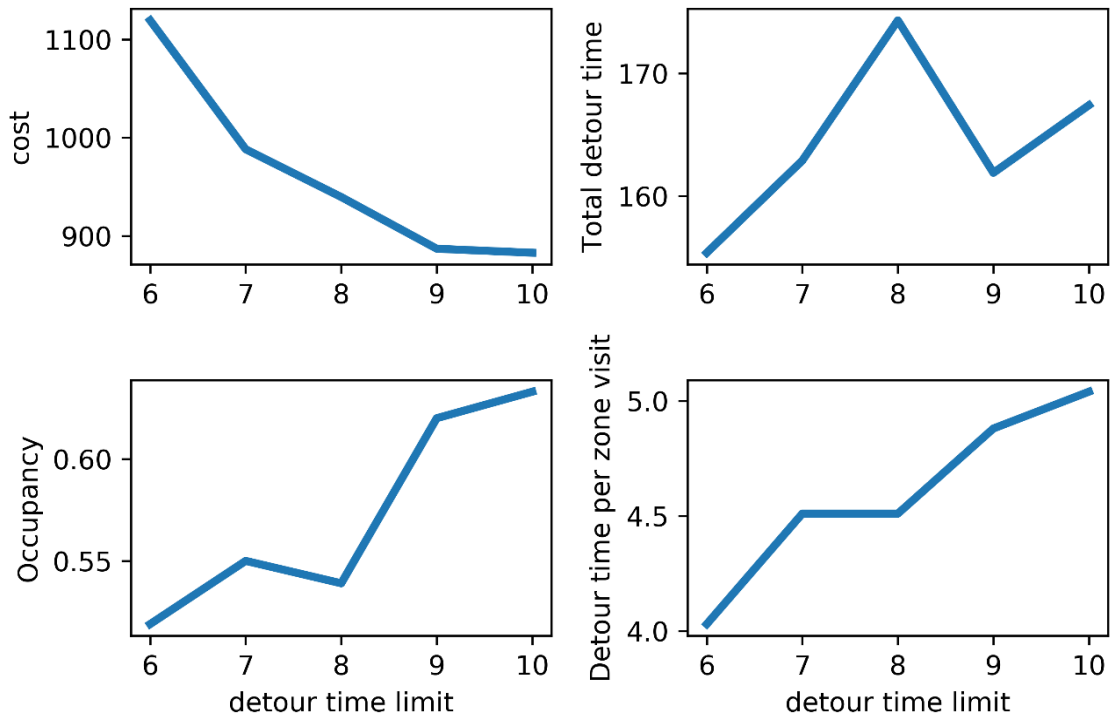

Figure 8. Total cost, vehicle occupancy and detour time versus detour time limit

Firstly, as the detour time constraint is relaxed, the total cost decreases. As Scenario 5 has the lowest detour time limit, the average ad hoc cost is significantly higher than that in other scenarios, because regular service is bounded by the lower detour time limit and therefore a lot of service requests can only be served by ad hoc services. As longer detour time is allowed, the service quality is worsened in terms of detour time. The detour time per zone visit increases from 4.03 minutes in Scenario 5 to 5.04 minutes in Scenario 8. The detour times are positively correlated with detour time limits, as some passengers that are not served due to the detour time constraint will be served after the constraint is relaxed, and therefore occupancies are also increased. The occupancy also increased due to a slight decrease of vehicle used. Similar to the sensitivity analysis on vehicle capacity, the total detour time is non-monotonic as Scenario 7 uses signicantly fewer flexible buses, as a result of relaxing the detour time limit.

To conclude, this example shows that 10-seaters are preferred. Also, a trade-off between service quality and total cost is observed through the sensitivity analysis on the maximum allowable detour time. Relaxation of the detour time limit leads to longer detour time per zone and higher occupancy. The regular service design, however, is subjected to other factors such as demand characteristics. To sum up, the proposed formulation is capable of planning vehicle deployment with different vehicle capacities and detour time limits.

### 3.4 Results and computation time for larger instance

As the number of passengers increases, more variables are introduced to **P2,** which increases the problem size and potentially increases the solution time. Following the setting of Scenario 3, this example evaluates the performance of the flexible bus system and computation time of the service reliability-based gradient solution approach based on different demand volume level. Two scenarios that multiply the parameter $\mu$ of demand volume distribution by 1.2 and 1.5 respectively are simulated. The detailed result is shown in Table 7.

Table 7. Output parameters under different factor of service request

|  | Scenario 3 | Scenario 9 | Scenario 10 |
|---|---|---|---|
| Demand volume multiplier | **1** | **1.2** | **1.5** |
| Total cost (HK$) | 939.6 | 1164.8 | 1388.7 |
| Vehicle operating cost $C_f$ (HK$) | 810 | 730 | 1130 |
| Ad hoc cost (HK$) | 129.6 | 434.8 | 258.7 |
| Average volume reliability $\bar{\rho}^I$ | 0.4789 | 0.2304 | 0.5169 |
| Average detour reliability $\bar{\rho}^{II}$ | 0.5 | 0.0421 | 0.5 |
| Vehicle used | 18 | 16 | 26 |
| Vehicle occupancy | 53.9% | 63.9% | 56.5% |
| Total detour time (mins) | 174.3 | 187.1 | 260.5 |
| Detour time per zone visit (mins) | 4.51 | 5.09 | 4.75 |
| Computation time(s) | 1984 | 11492 | 26621 |
| Number of iterations | 2 | 8 | 3 |
| Average time per iteration (s) | 992 | 1437 | 8873 |

It is observed that the total cost and total detour time increase with the number of service requests. The economics of scale are not significant as the total cost increases by 48% while the demand volume increases by approximately 50%. Computationally, the average time per iteration increases rapidly with the service request factor. More precisely, the average time per iteration increases about 900% with a 50% increase in demand.

### 3.5 Application for Chengdu, China, with real data

Finally, to illustrate the formulation for a real scenario, we apply it to Chengdu, China, by using the empirical service request data from Didi Chuxing from 1st November to 30th November 2016. The flexible bus service is planned based on the data from 9:00am to 9:15am as representative of the morning peak. The area within the second ring road is divided into nine zones as shown in Figure 9. The amount and detour times of service requests are derived from the real data, but the volume is scaled down to 20%, as detailed in Appendix G. The cost to travel from a zone to an adjacent zone with shared border (e.g. from zone A to zone D) is RMB¥5, while to an adjacent zone with shared corner only (e.g. from zone A to zone E) is RMB¥7. To cope with the larger zone size, the maximum allowable detour time within a zone is 15 minutes, the vehicle capacity is 10 passengers.

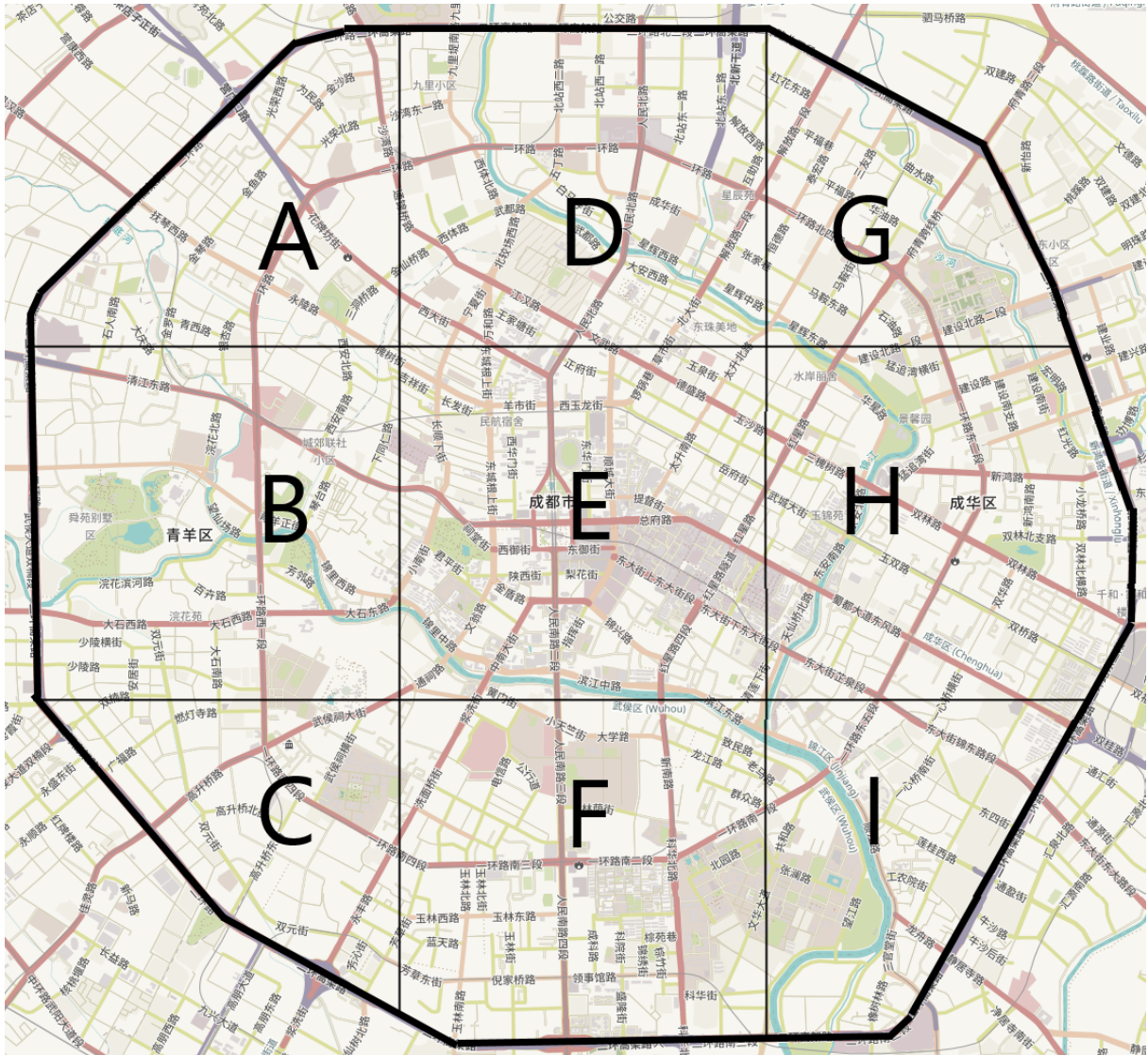

Figure 9. The study area in Chengdu, zone divisions and labels (Source: Map data OpenStreetMap contributors, CC BY-SA.)

The solution shows that 45 vehicles are used. Specifically, each of AE, BC, BD, BEH, CB, CE, CFI, DBC, EA, ED, EG, EH, FBA, FB, FH, GH, HEB, HEC, HD, HF, HG, IE, IF, IH, CFH and BFI is assigned a vehicle, two vehicles are assigned to route DEF, DHI, EB, FC and FED, and three vehicles are assigned to EC, EF and EI. The total cost is RMB¥496.11, where the regular service cost and ad hoc service cost are RMB¥343.83 and RMB¥152.28, respectively. 88.2% of the passengers are served by the flexible bus, the average number of service locations visited per zone is 4.20, and the average detour time per zone is 10.93 minutes. As the average demand volume is large, the computation time is 165 hours in 6 iterations.

## 4. Conclusion

In this paper, we proposed a zonal-based flexible bus operation strategy. Demands were clustered geographically into zonal OD pairs and number of passengers. The volume and spatial stochastic variations of demand were estimated by historical record. A two-stage stochastic problem with recourse was formulated to plan the flexible bus routes. Volume reliability measure and detour time reliability measure were introduced to divide the otherwise intractable two-stage stochastic problem into two phases. In phase-1, the zonal visit sequences of vehicles were optimized to serve customer demand up to a certain level of reliability, while satisfying the vehicle capacity and maximum allowable detour time. Upon the realization of demand, the phase-2 optimization matched service requests to vehicles. A service-reliability based gradient solution approach was then implemented to solve the problem by finding optimal reliability measures that minimized the sum of fixed operating cost and ad hoc service cost incurred for customers who cannot be served by regular service. Given the gradient direction, the step sizes were calculated by either a procedure or the Adam method. In our numerical experiments, this hybrid method converged more consistently than either method. It was proved that any optimal solutions obtained in the original two-stage stochastic problem with recourse **P0** can be obtained by the SR-based gradient solution approach **P1**-**P2**, and vice versa. The numerical studies illustrated that the SR-based gradient solution approach consistently produce good solutions under different probability distributions. In a case study, it was demonstrated that vehicles with capacity 10 were preferred, based on the trade-off among the total operating cost, detour time, and occupancy. The results provided insights for operators to decide the routing, vehicle capacity and fleet size based on the aforementioned criteria. The result for the numerical example with a larger demand showed that the computation time increased rapidly with the problem size. Applying the model to Chengdu, China, based on real sample data required over 165 hours of computation time. How to develop a more efficient solution approach for a large demand is an important topic for further studies.

The proposed model provides a novel way to formulate and optimize a flexible bus system. The zonal visit sequences can either be treated as traditional long-term strategic planning that is optimized when the demand pattern changes substantially, or short-term planning that is responsive to demand changes due to stochastic demand characteristics. Operationally, the actual route is determined by solving the passenger-to-vehicle assignment problem in real-time, drivers then have the flexiblility on routing as long as all the required pickup or drop off points are visited. Currently, the objective function is only based on the benefit of the mobility service providers, and the benefits of other stakeholders are not considered. Future researches can be conducted to include different stakeholders' benefits, such as user costs, waiting time and profit. The other possible extensions include improving existing or developing alternative solution methods to solve larger problem sizes more efficiently, incorporating time windows for pick-up and drop-off, and optimizing for dynamic demand realization.

## Appendix A. Illustration of converting matrix $\mathbf{B}_p$, detour correlation matrix $\mathbf{T}_z$ and capacity constraint

To illustrate detour constraint (5) and vehicle capacity constraint (7), we consider demand realization with 4 different service requests, with the detailed information shown in Table 8. The OD of passengers onboard in each portion of route ABC is shown in Figure 10.

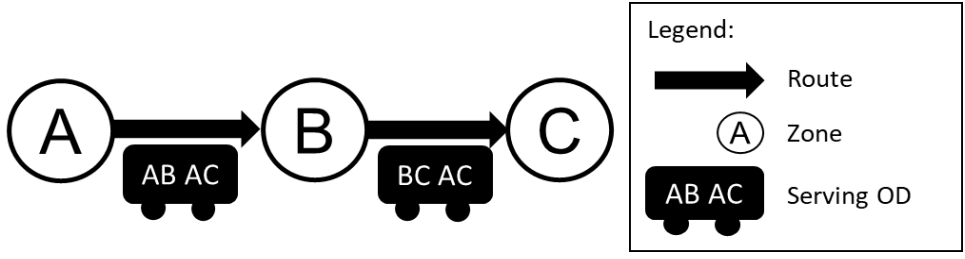

Figure 10. The OD of passengers onboard from A to B and from B to C.

Table 8. Detailed information of 4 service requests

| Service request no. $d$ | Number of passengers $n_d$ | Origin (detour time $\tau_{d+}$, minutes) | Destination (detour time $\tau_{d-}$, minutes) | Ad hoc cost $c_d$ (HK$) |
|---|---|---|---|---|
| 1 | 2 | A (1) | C (2) | 6 |
| 2 | 3 | A (2) | C (1) | 6 |
| 3 | 2 | A (3) | B (2) | 3 |
| 4 | 1 | B (2) | C (2) | 2 |

From the above table, we have $\alpha_{1+}^A = \alpha_{2+}^A = \alpha_{3+}^A = \alpha_{4+}^B = \alpha_{1-}^C = \alpha_{2-}^C = \alpha_{3-}^B = \alpha_{4-}^C = 1$ indicates the origins and destinations of each service request, and the other $\alpha_{d+(-)}^z$ values are zero for every zone $z$ and service request $d$.

Moreover, the detour time matrices are given by,

$$\mathbf{T}_A = \begin{bmatrix} 1 & -0.2 & -0.2 & 0 \\ -0.2 & 2 & -0.5 & 0 \\ -0.2 & -0.5 & 3 & 0 \\ 0 & 0 & 0 & 0 \end{bmatrix}, \mathbf{T}_B = \begin{bmatrix} 0 & 0 & 0 & 0 \\ 0 & 0 & 0 & 0 \\ 0 & 0 & 2 & -0.5 \\ 0 & 0 & -0.5 & 2 \end{bmatrix}, \mathbf{T}_C = \begin{bmatrix} 2 & -0.2 & 0 & -0.5 \\ -0.2 & 1 & 0 & -0.25 \\ 0 & 0 & 0 & 0 \\ -0.5 & -0.25 & 0 & 2 \end{bmatrix}$$

The diagonal of matrix $\mathbf{T}_z$ is the detour time required to serve respective service requests. The other elements are the reduction of detour time by picking up or dropping off more than one service requests at a zone simultaneously. This matrix satisfies condition (13) as the absolute value of the sum of the detour time reduction in every row is smaller than half of the additional detour time. For example, in the first row of $\mathbf{T}_A$, the absolute value of sum of the detour time reduction is 0.4, which is smaller than half of the additional detour time 1 at that row. Thus, adding any service requests to a vehicle will only increase the detour time. With the correlation matrices, the detour constraint (5) is stated as,

$$\begin{aligned} w_{1v} + 2w_{2v} + 3w_{3v} \quad -0.4w_{1v}w_{2v} - 0.4w_{1v}w_{3v} - w_{2v}w_{3v} &\leq \bar{t}_A, \quad \forall v \in V \\ +2w_{3v} + 2w_{4v} - w_{3v}w_{4v} &\leq \bar{t}_B, \quad \forall v \in V \quad (44) \\ 2w_{1v} + 1w_{2v} \quad +2w_{4v} - 0.4w_{1v}w_{2v} - w_{1v}w_{4v} - 0.5w_{2v}w_{4v} &\leq \bar{t}_C, \quad \forall v \in V \end{aligned}$$

where $w_{dv}$ is the binary variable indicating whether request $d$ is served by vehicle $v$. Equation (6) calculates the number of passengers for each OD pair served by every vehicle $v \in V$ as follows.

$$\zeta_v^{AC} = 2w_{1v} + 3w_{2v}$$
$$\zeta_v^{AB} = 2w_{3v} \qquad (45)$$
$$\zeta_v^{BC} = w_{4v}$$

We use route $p = \text{ABC}$, i.e. from zone A to B, then from B to C, to illustrate the converting matrix $\mathbf{B}_p$, matrix $\mathbf{B}_{\text{ABC}}$ is given by,

$$\mathbf{B}_{\text{ABC}} = \begin{bmatrix} 1(=b_A^{AB}) & 0(=b_A^{BC}) & 1(=b_A^{AC}) & M_1 & M_1 & M_1 \\ 0(=b_B^{AB}) & 1(=b_B^{BC}) & 1(=b_B^{AC}) & M_1 & M_1 & M_1 \end{bmatrix} \qquad (46)$$

The rows represent the zones where passengers can be picked up; in this case, zone A and B, correspond to the first and the second row respectively. The columns show all the OD pairs in the following order: AB, BC, AC, CB, BA, and CA. Precisely, the passengers of OD pairs AB and AC are onboard in zone A, with corresponding $b_A^{AB}$ and $b_A^{AC}$ equal to 1 in the first row. $b_B^{AB}$ is zero because passengers of OD pair AB will not onboard when the vehicle leaves zone B. Moreover, the entries for OD pairs CB, BA, and CA are represented by a large positive number $M_1$, which ensures that every vehicle of route ABC does not serve any service requests from these OD pairs that route ABC does not traverse.

With the converting matrix $\mathbf{B}_{\text{ABC}}$, if $x_{\text{ABC},v} = 1$, the capacity constraint (7) for vehicle $v$ can be expressed as:

$$\zeta_v^{AB} + \zeta_v^{AC} + M_1\zeta_v^{CB} + M_1\zeta_v^{BA} + M_1\zeta_v^{CA} \leq cap \quad \text{for zone A}$$
$$\zeta_v^{BC} + \zeta_v^{AC} + M_1\zeta_v^{CB} + M_1\zeta_v^{BA} + M_1\zeta_v^{CA} \leq cap \quad \text{for zone B} \qquad (47)$$

Again the large positive number $M_1$ ensures that vehicle $v$ of route ABC will not serve any passengers other than the OD pairs in the route ABC, i.e., $\zeta_v^{CB}$, $\zeta_v^{BA}$ and $\zeta_v^{CA}$ must be zero if $x_{\text{ABC},v}$ is equal to 1.

If there are two vehicles in the fleet, i.e., $|V| = 2$, $v = 1$ for the first vehicle and $v = 2$ for the another, and cost of route ABC $c_{\text{ABC}}^1 = \text{HK\$}10$, combining constraints (44), (45) and (47), the optimal solution is to deploy one vehicle to route ABC if the maximum allowable detour time of each zone $\bar{t}_A, \bar{t}_B, \bar{t}_C \geq 4.2$ and the vehicle capacity $cap \geq 7$, and the total operating cost is HK\$10. If $\bar{t}_A < 4.2$ and $cap \geq 6$, the optimal solution is to deploy one vehicle to route ABC to serve request 1, 2, 4 and allocate ad hoc service for request 3, i.e., $x_{\text{ABC},1} = 1, x_{\text{ABC},2} = 0$, $\{w_{1,1}, w_{2,1}, w_{3,1}, w_{4,1}\} = \{1,1,0,1\}$ and the total operating cost, including the HK\$3 additional ad hoc service cost is HK\$13. There is a difference since the sum of detour time in zone A is 4.2 if we serve service requests 1, 2 and 3 simultaneously. Note that as there are multiple optimal solutions by swapping routes of vehicles, for instance, the objective function above is indifferent with $x_{\text{ABC},1} = 0, x_{\text{ABC},2} = 1$, $\{w_{1,2}, w_{2,2}, w_{3,2}, w_{4,2}\} = \{1,1,0,1\}$.

## Appendix B. Proof of Proposition 1

**Proposition 1.** *Any feasible solution of **P1-P2** is also feasible to the original two-stage stochastic formulation **P0**.*

**Proof.** We consider a solution (**X,Y,W**) after solving **P1** and **P2**. Suppose **X,Y** is a solution of **P1** for a given service reliability $\boldsymbol{\rho}$, and **W** is a solution of **P2** for demand realization $\kappa$. Constraints (2) and (3) in **P0** are satisfied as they are the same as (17) and (25) in **P1**. Moreover, constraints (5)–(10) are satisfied as **W** is feasible in **P2** and (10) is equal to (37).

In short, for any feasible solution (**X,Y,W**) of **P1** and **P2**, constraints (2), (3) and (5)-(10) are fulfilled, so it is feasible in the two-stage stochastic formulation **P0**. $\square$

## Appendix C. Derivation and Proof of Proposition 2 and Proposition 3

To show that optimal solutions of **P0** can always be obtained by solving **P1** for some reliability measures, with the condition that there is one demand category per OD, we show that all solutions of **P0** are reproducible from **P1** by setting appropriate reliability measures.

To facilitate the proof below, some notation and definitions are introduced. Firstly, by assumption B1, there is always one shortest path (defined as *direct route*) starting exactly from zone I and ending exactly in zone J for every OD pair $(I,J) \in \Omega$, denoted by $P_{IJ}$, which has an operating cost denoted by $c_{IJ}^1$. Secondly, we introduce the notion of *active* detour time constraint. Given any demand category of OD pair IJ and any vehicle $v \in V$ with route $P_{IJ}$, the capacity constraint (21) ensures that the number of service requests of type $e$ picked up by vehicle $v$, i.e., $y_{ev}$, is bounded above by $\left\lfloor \frac{cap}{n_e} \right\rfloor$, $n_e$ is the number of passengers in $e$. Moreover, from the detour time equations (14), (15) and (23), $y_{ev}$ is also bound by $y_{ev}^{\max,time} = \sup\{\eta \mid \eta \in \mathbb{Z} : 2\tilde{\tau}_I(\eta) + (\eta-1)\tau_I^{II} \leq \bar{t}_I \land 2\tilde{\tau}_J(\eta) + (\eta-1)\tau_J^{II} \leq \bar{t}_J\}$. The detour constraint is said to be *active* if $\left\lfloor \frac{cap}{n_e} \right\rfloor > y_{ev}^{\max,time}$, which means that the maximum number of passengers picked up is constrained by the maximum detour bound, but not by the capacity constraint. In the Lemma 1 and Proposition 2 below, by setting appropriate detour reliabilities, the detour constraint is not active, thus the number of requests of service request $e$, with OD pair IJ, picked up by any vehicle serving route $P_{IJ}$ is bounded by,

$$y_{ev} \leq \left\lfloor \frac{cap}{n_e} \right\rfloor \qquad (48)$$

**Lemma 1.** *Given that only the demand category of one passenger exists for each OD pair, if the detour constraints are inactive, and vehicle capacity cap is a factor of demand volume $\delta_e, \forall e \in E$, serving all demands by their respective direct routes minimizes the operating cost. (i.e. better than or equal to other combinations).*

**Proof.** Define $\sigma_{IJ}^{AB}$ as the cost of providing a space for one passenger from zone I to zone J

by route $P_{AB}$. The spaces may or may not be occupied by a passenger, but the number of spaces provided must be higher than the number of passengers of that OD served. Based on the objective function (16), assigning a vehicle to route $P_{IJ}$ costs $c_{IJ}^1$. On the other hand, equation (48) states that it can provide space for at most *cap* service requests of OD pair IJ as all demand categories have one passenger ($n_e = 1$). Therefore, for any pair of zone IJ, we can express the cost of space provided by its direct route as,

$$\sigma_{IJ}^{IJ} = \frac{c_{IJ}^1}{cap} \tag{49}$$

Apart from the direct route $P_{IJ}$, service requests of OD pair IJ may be served by vehicles with three types of routes traversing zone I and zone J: (1) route starts in zone I, passes through zone J and ends in some other zones; (2) route starts somewhere neither I nor J, passes through zone I and ends in zone J; and (3) route starts somewhere neither I nor J, passes through zone I and zone J respectively, and ends elsewhere. We will show that serving the service requests by these routes does not reduce the operating cost.

In case (3), there is a route $P_{AB}$ starting at specific zone A that is neither I nor J, passes through zone I and then zone J, and ends at specific zone B that is neither I nor J. Following the vehicle capacity constraint (21), if a vehicle is serving $n$ service requests from zone I to zone J, it can serve at most $cap - n$ service requests from A to B and $n$ service requests from A to I and from J to B, as illustrated in Figure 11. Serving one customer from I to J will occupy one space for customer from A to B, vice versa. The space from A to I, and J to B obtained by serving customers from I to J can either be filled in by customers from A to I or left vacant. Therefore, for any $n \leq cap$, we have the following relationship between $\sigma_{AI}^{AB}$, $\sigma_{IJ}^{AB}$, $\sigma_{JB}^{AB}$ and $\sigma_{AB}^{AB}$.

$$n\sigma_{AI}^{AB} + n\sigma_{IJ}^{AB} + n\sigma_{JB}^{AB} + (cap - n)\sigma_{AB}^{AB} = c_{AB} \tag{50}$$

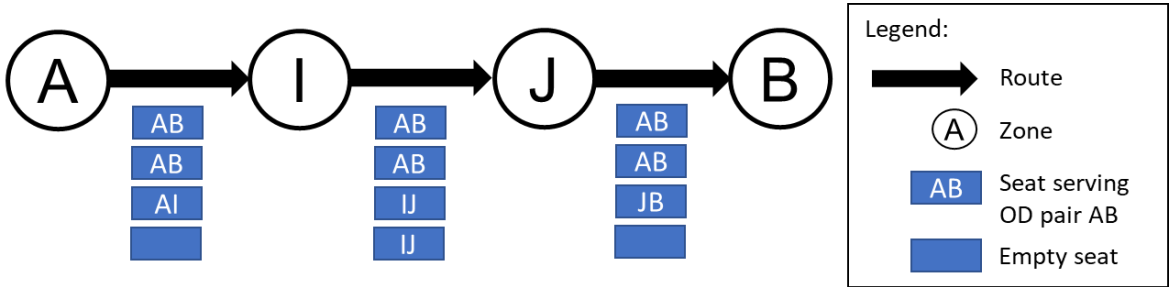

Figure 11. Seat allocation of route AIJB.

From assumption B2, we have $c_{AB} = c_{AI} + c_{IJ} + c_{JB}$, by substitution to (50),

$$n\sigma_{AI}^{AB} + n\sigma_{IJ}^{AB} + n\sigma_{JB}^{AB} + (cap - n)\sigma_{AB}^{AB} = \frac{cap - n}{cap}c_{AB} + \frac{n}{cap}c_{AI} + \frac{n}{cap}c_{IJ} + \frac{n}{cap}c_{JB} \tag{51}$$

Since $P_{AB}$ is the (shortest path) direct route to serve demand from A to B, from (49), we have $\sigma_{AB}^{AB} = \frac{c_{AB}}{cap}$, by substitution to (51) and division by $n$,

$$\sigma_{AI}^{AB} + \sigma_{IJ}^{AB} + \sigma_{JB}^{AB} = \frac{1}{cap}(c_{AI} + c_{IJ} + c_{JB}) \tag{52}$$

Equation (52) establishes the relationship between costs of providing space $\sigma_{AI}^{AB}$, $\sigma_{IJ}^{AB}$, $\sigma_{JB}^{AB}$ by route $P_{AB}$ and route operating costs $c_{AI}$, $c_{IJ}$ and $c_{JB}$. It is equal to the cost of providing space with the direct route, as the cost of providing a space for passengers of OD pair AI, IJ and JB by direct route are $\sigma_{AI}^{AI} = \frac{c_{AI}}{cap}$, $\sigma_{IJ}^{IJ} = \frac{c_{IJ}}{cap}$ and $\sigma_{JB}^{JB} = \frac{c_{JB}}{cap}$ respectively. As vehicle capacity is a factor of the demand volume, serving every service request by the respective direct routes left no vacant space. The other allocation of service requests $P_{AB}, P_{AI}, P_{IJ}$ and $P_{JB}$ that also waste no seats have the same cost as shown in the above as all spaces are used up. The same conclusion can be drawn for case (1) and case (2) mentioned above by either removing terms with AI or terms with JB.

To conclude, as the vehicle capacity is a factor of demand volume, serving the service requests by direct route leaves no empty seats, thus minimizes the cost of providing space. Therefore, the operating cost is minimized by serving the service requests by the direct route. □

**Proposition 2**. *For problems with the stochastic volume described by a probability distribution for any integer greater than 0, and the stochastic detour time described by a probability distribution over the range* $[0, +\infty)$, *if there exists only one demand category with one passenger per OD, any optimal solution of the original 2-stage stochastic formulation can be generated by the SR-based stochastic formulation proposed.*

**Proof.** The general idea of the proof is that for any optimal solution $(\mathbf{X}^*, \mathbf{W}^*)$ in **P0**, there exists some $\boldsymbol{\rho}^I$ and $\boldsymbol{\rho}^{II}$ such that the resultant solution from **P1** is $\mathbf{X}^*$.

Suppose **P0** has an optimal solution $(\mathbf{X}^*, \mathbf{W}^*)$, we can calculate $m_p^*$, the number of vehicles assigned to route $p$ (i.e. $m_p^* = \sum_{v \in V} x_{pv}^*$). For every route $p$ where $m_p^* > 0$, we find the corresponding demand category $e \in E$ with identical origin zone I and destination zone J with route $p$. Then, for that demand category $e$, with the condition that the demand volume equal to the amount of vehicle spaces, $\delta_e = m_p^* cap$, and detour constraints are inactive (e.g. A reliability that makes every $\tau_z = 0$), we can always obtain $\rho_e^I, \rho_z^{II}$ by inverting the probability distributions given in Equation (18) and Equation (22). Combining reliabilities from all demand categories, we have $\boldsymbol{\rho}^I$ and $\boldsymbol{\rho}^{II}$. By defining reliability as $\boldsymbol{\rho}^I$ and $\boldsymbol{\rho}^{II}$, we have $cap$ as a factor of $\delta_e$, and detour constraints are not active since they do not further restrict the number of service requests can be served. By Lemma 1, in the above condition, serving all service request of category $e$ by vehicles with direct route $p$ yields the least cost. Therefore, $\mathbf{X}^*$ is an optimal solution to **P1**, given $\delta_e = m_p^* cap$, $\tau_z^{II}$ and by designating $\boldsymbol{\rho}^I$ and $\boldsymbol{\rho}^{II}$. In short, for any optimal solution $\mathbf{X}^*$ of the original two-stage formulation, it is always possible to identify some $\boldsymbol{\rho}^I$ and $\boldsymbol{\rho}^{II}$ in **P1** to reproduce $\mathbf{X}^*$. □

**Proposition 3.** *For the same problem in Proposition 2, with multiple demand categories per*

OD, including single passenger service requests, any optimal solution of the original 2-stage stochastic formulation can be generated by the SR-based stochastic formulation.

**Proof.** This proposition extends Proposition 2 to multiple types of demand per OD. Given an optimal solution $(\mathbf{X}^*, \mathbf{W}^*)$ in **P0**, despite there are multiple types of demand per OD, we can always generate $\mathbf{X}^*$ by varying the volume reliabilities of those single person requests and set the reliability measures of multiple people requests as 0.

First, set the reliability $\rho_e^I = 0$ for demand category $e$ such that $n_e > 1$, so the service provision for these demand categories is 0 by (18). Therefore, the corresponding $y_{ev}$ is always zero, and the problem is reduced to one type of demand with one passenger per OD by condition (20) and (23). Follow Proposition 2, for each route $p$ with vehicle assigned to, that OD is $(I,J) \in \Omega$, we find a demand category $e$ with OD pair (I,J) and one passenger (i.e. $n_e = 1$), and set each $\rho_e^I$, $\rho_z^{II}$ such that $\delta_e = m_p^* cap$ and $\tau_z^{II} = 0$ by inverting the probability distributions. □

## Appendix D. Algorithms for step size determination

For every demand category $e$, there is a set of route $P_e \subseteq P$ that can serve that demand category. The spaces left for vehicles assigned to those routes are checked, and the number of additional service requests from demand category $e$ can be served is calculated. First, we consider the capacity constraint (21), which uses $\mathbf{B}_p$ to map the number of passengers per OD $\zeta_v$ to a vector of passengers onboard when leaving each zone along the route. We express $\mathbf{B}_p \zeta_v$ by the following vector,

$$\mathbf{B}_p \zeta_v = \begin{pmatrix} \mu_{1v} & \mu_{2v} & \ldots & \mu_{m_p-1,v} \end{pmatrix}^T \qquad (53)$$

where $\mu_{iv}$ is the number of passengers on vehicle $v$ in the $i$-th zone along route $p$. The vector has only $m_p - 1$ entries as we do not need to consider the capacity constraint for the end zone. The capacity constraint (21) limits $\mu_{iv} \leq cap, \forall i, \forall v \in V$.

If $\mathbf{X}^*$ does not change, the maximum number of additional service requests from demand category $e$ can be served by vehicle $v$, denoted as $\epsilon_{ev}^I$, is bounded by the minimum slack capacity along the route, written as,

$$\epsilon_{ev}^I \leq \frac{\min_{i \in Z_{ev}}(cap - \mu_{iv})}{n_e} \qquad (54)$$

$n_e$ is the number of passengers of demand category $e$, $Z_{ev} \subseteq Z$ is the set of visited zones of vehicle $v$ (ending zone excluded) for any service request of $e$. Next, we consider the detour time constraints. $\epsilon_{ev}^I$ can be calculated by rearranging the terms in the detour constraint (14), (15) and (23), where zone R and zone S are the origin and destination of demand category $e$. Therefore, we have equation (40) used in Section 2.4.5.1.

$$\epsilon_{ev}^{\mathrm{I}} = \left\lfloor \min\left[ \frac{\min_{i \in Z_{ev}}(cap - \mu_{iv})}{n_e}, \frac{\bar{t}_{\mathrm{R}} - t_{v\mathrm{R}}}{\tau_{\mathrm{R}}^{\mathrm{II}}}, \frac{\bar{t}_{\mathrm{S}} - t_{v\mathrm{S}}}{\tau_{\mathrm{S}}^{\mathrm{II}}} \right] \right\rfloor \qquad (40)$$

## Appendix E. Proof of Proposition 4 and Proposition 5

**Proposition 4.** The objective value of **P1** will not change if the volume of demand category $e$, $\delta_e^{\mathrm{I}}$, is increased by any $0 \leq \Delta \delta_e^{\mathrm{I}} \leq \epsilon_e^{\mathrm{I}}$, where $\epsilon_e^{\mathrm{I}}$ is the maximum demand volume increment given by Algorithm 1.

**Proof.** We prove by showing that for each demand category $e$ considered separately, all the constraint will not be violated by increasing $\delta_e^{\mathrm{I}}$ by any value $\Delta \delta_e^{\mathrm{I}}$ less than $\epsilon_e^{\mathrm{I}}$.

Based on $\epsilon_{ev}^{\mathrm{I}}$ specified for each vehicle $v$ in Step 2.1.1 in Algorithm 1, we define the increment of requests picked up by vehicle $v$ as $\Delta \delta_{ev}^{\mathrm{I}}$, that satisfies,

$$\sum_{v \in V} \Delta \delta_{ev}^{\mathrm{I}} = \Delta \delta_e^{\mathrm{I}} \leq \epsilon_e^{\mathrm{I}} = \sum_{v \in V} \epsilon_{ev}^{\mathrm{I}}$$

Furthermore, from Step 2.1.1 of Algorithm 1, we have,

$$\Delta \delta_{ev}^{\mathrm{I}} \leq \epsilon_{ev}^{\mathrm{I}} \leq \frac{cap - \mu_{iv}}{n_e}, \forall i \in Z_{ev}, \forall v \in V$$

Rearranging the term, we get

$$\mu_{iv} + \Delta \delta_{ev}^{\mathrm{I}} n_e \leq cap, \forall i \in Z_{ev}, \forall v \in V$$

As the passengers of demand category $e$ are only onboard when leaving zones in $Z_{ev}$, we retain the inequality $\mathbf{B}_p \boldsymbol{\zeta}_v \leq cap \mathbf{1}_{m_p - 1}$ for $x_{pv} = 1$. Therefore, the constraint (21) is fulfilled.

On the other hand, we have $\Delta \delta_{ev}^{\mathrm{I}} \leq \epsilon_{ev}^{\mathrm{I}} \leq \frac{\bar{t}_{\mathrm{R}} - t_{v\mathrm{R}}}{\tau_{\mathrm{R}}^{\mathrm{II}}}$ at origin zone R of $e$. Rearranging the term, we have,

$$\begin{aligned}
\bar{t}_{\mathrm{R}} &\geq t_{v\mathrm{R}}\left(\tilde{y}_{v\mathrm{R}}^*\right) - 2\tilde{\tau}_{\mathrm{R}}\left(\tilde{y}_{v\mathrm{R}}^*\right) + \epsilon_{ev}^{\mathrm{I}} \tau_{\mathrm{R}}^{\mathrm{II}} + 2\tilde{\tau}_{\mathrm{R}}\left(\tilde{y}_{v\mathrm{R}}^*\right) \\
&= 2\tilde{\tau}_{\mathrm{R}}\left(\tilde{y}_{v\mathrm{R}}^*\right) + \left(\epsilon_{ev}^{\mathrm{I}} + \tilde{y}_{v\mathrm{R}}^*\right)\tau_{\mathrm{R}}^{\mathrm{II}} \\
&\geq 2\tilde{\tau}_{\mathrm{R}}\left(\tilde{y}_{v\mathrm{R}}^* + \epsilon_{ev}^{\mathrm{I}}\right) + \left(\tilde{y}_{v\mathrm{R}}^* + \epsilon_{ev}^{\mathrm{I}}\right)\tau_{\mathrm{R}}^{\mathrm{II}} \\
&= t_{v\mathrm{R}}\left(\epsilon_{ev}^{\mathrm{I}} + \tilde{y}_{v\mathrm{R}}^*\right)
\end{aligned} \qquad (55)$$

Therefore, detour constraint (23) is fulfilled at origin zone R. Similarly, $\Delta \delta_e^{\mathrm{I}} \leq \epsilon_{ev}^{\mathrm{I}} \leq \frac{\bar{t}_{\mathrm{S}} - t_{v\mathrm{S}}}{\tau_{\mathrm{S}}^{\mathrm{II}}}$ fulfills detour constraint (23) at destination zone S. Therefore, every constraint in **P1** is fulfilled and the objective value will not be changed. $\square$

**Proposition 5.** Given passenger assignments to vehicle $\mathbf{Y}^*$, for any demand category $e$, the objective value of **P1** does not change if the travel time between requests of zone $z$ is increased by any $\Delta \tau_z^{\mathrm{II}} \leq \epsilon_z^{\mathrm{II}}$

**Proof.** For every vehicle $v$ and every zone z, we have
$$\Delta\tau_z^{II} \leq \epsilon_z^{II} \leq (\overline{t}_z - t_{vz})/\tilde{y}_{vz}^*, \forall v \in V$$
Rearranging the term, we have $t_{vz} + \tilde{y}_{vz}^*\Delta\tau_z^{II} \leq \overline{t}_z, \forall v \in V$. As the new detour time between two requests in zone $z$ will become $\tau_z^{II} + \Delta\tau_z^{II}$, the detour constraint (23) is fulfilled. □

## Appendix F: Decomposition of the problem

Without loss of generality, we can decompose the problem to reduce the computational time if there exist a set of service requests $D_i \subseteq D$ or demand category $E_i \subseteq E$ and a set of routes $P_i \subseteq P$, where each service request $d \in D_i$ or demand category $e \in E_i$ can only be served by routes $p \in P_i$, and each route $p \in P_i$ can only serve service requests in $D_i$ or demand categories in $E_i$. Then, as the set of routes and set of service request neither affecting nor being affected by other routes and demand categories, **P1** and **P2** can be solved for a smaller set. For example, in the six-zone scenario depicted in Figure 12, if there is demand from OD pairs AB, AE, CD and BF, and the possible routes are AB, ACDE and BCDF, **P1** can be separated into two parts. The first part is OD pair AB with possible route AB, the second part is of OD pairs AE, CD and BF, and possible routes ACDE and BCDF. Formally, $P_0 = \{AB\}$, $D_0 = \{d \in D \mid \alpha_{d+}^A \alpha_{d-}^B = 1\}$, $P_1 = \{ACDE, BCDF\}$ and $D_1 = \{d \in D \mid \alpha_{d+}^A \alpha_{d-}^E + \alpha_{d+}^C \alpha_{d-}^D + \alpha_{d+}^B \alpha_{d-}^F = 1\}$. This decomposition by OD can reduce computational time significantly as both **P1** and **P2** are NP-hard integer programming problem which the computational times increase exponentially with the size of demand and size of routes.

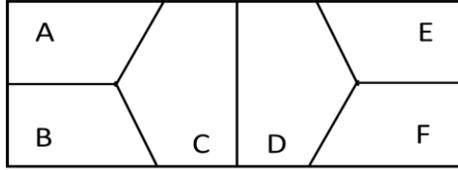

Figure 12. The six-zone scenario

## Appendix G: Usage of real data: a case study in Chengdu

In this study, the data from Tuesdays, Wednesdays, and Thursdays between 1st and 30th of November from 9:00am to 9:15am are used. As the coordinate system of the original data is at GCJ-02 coordinate system, it is first transformed to a Cartesian coordinate system in meter unit. The Euclidean distance is then calculated based on the transformed coordinate system between zone centroid and the location. After that, the detour time is obtained by multiplying the distance by 0.003min/m, which came from a road detour ratio of 1.25 given in the study of Yang et al. (2018) and an assumption that the average speed is 25km/h in Chengdu city area.

The demand volume distribution adapted the empirical distribution of number of ride requests between zones in the study period. The function $\tilde{\tau}_z$ between the boundary of a zone and the closest service request is inferred by the following algorithm. As the zone is divided by squares, the boundary can be represented by 2 horizontal lines and 2 vertical lines which form a square between them.

**Algorithm 3. Approximate the detour time function between boundary and the closest**

service request from the data

**Input:** Sets of origin and destination location at each zone, the boundary of each zone $(x^z_{max}, x^z_{min}, y^z_{max}, y^z_{min})$

**Output:** Approximation values of $\tilde{\tau}_z(i)$ for each zone $z \in Z$ and meaningful $i$.

**Algorithm:**
  **For each** zone $z \in Z$
    **For each** number of service request considered $i$
      Conduct the below calculation 1000 times
        Randomly sample $i$ locations in zone $z$ as $P$, each with x, y coordinate denoted by $(x, y)$
        Denote $P_X^{max}, P_X^{min}, P_Y^{max}, P_Y^{min}$ as the maximal x coordinates, minimal x coordinate, maximal y coordinates, minimal y coordinates of locations in set $P$
        Distance $d \leftarrow \dfrac{P_X^{max} - x^z_{max} + x^z_{min} - P_X^{min} + P_Y^{max} - y^z_{max} + y^z_{min} - P_Y^{min}}{4}$
      Set the function value $\tilde{\tau}_z(i)$ by the average of $d$ times 0.003
    **End for**
  **End for**
**End**

Based on the data point approximated by the above algorithm, a parameter fitting for the exponential function $ae^{-bx} + c$ was conducted for each zone as $\tilde{\tau}_z(i)$ used in the phase-1 explained in Section 2.4.1. The result parameters are given in Table 9 below,

Table 9. The parameters fitted by zone

| Parameter | Zone | | | | | | | | |
|---|---|---|---|---|---|---|---|---|---|
| | A | B | C | D | E | F | G | H | I |
| a | 2.901 | 3.276 | 3.135 | 3.393 | 3.150 | 3.399 | 2.895 | 3.333 | 2.811 |
| b | 0.308 | 0.300 | 0.265 | 0.350 | 0.277 | 0.348 | 0.233 | 0.315 | 0.273 |
| c | 0.969 | 0.564 | 0.681 | 0.480 | 0.669 | 0.474 | 0.933 | 0.510 | 1.041 |

To generate a scenario based on historical data, for each zonal OD pair, the demand volume is generated by randomly choosing one of the historical volumes in the study period. The origin and destination locations are then randomly chosen from the historical ride requests with the same zonal OD without replacement. Next, the detour time $\tau^z_d$ for each zone $z$ to serve request $d$ are calculated from the distance between the coordinate and zone centroid, and the detour time reductions introduced in Section 2.3 are calculated from the proximity of the service request first given by,

$$\tau^{z-}_{db} = -\min(\tau^z_d, \tau^z_b) \times \left[ 1 - \frac{\sqrt{(x^z_d - x^z_b)^2 + (y^z_d - y^z_b)^2}}{3830.28} \right]$$

where $x^z_d$, $y^z_d$ are the x- and y-coordinate of service request location in zone $z$, and 3830.28 is the maximum Euclidean distance within a zone in meter scale, given the 3x3 zonal division. Finally, all detour time reduction $\tau^{z-}_{db}$ is normalized by a factor to ensure the sum of detour time reduction is less than the detour time, i.e. satisfying equation (13).

In short, this appendix explained how the data are chosen, calculation of demand volume,

detour time, detour time reduction, and the approximation of detour time between boundary and the closest service request from real data. The same methodology can be expanded to any set of ride request data with origin and destination location to conduct the ZBFBS analysis.

## Acknowledgements

The authors acknowledge Didi Chuxing providing ride request data for this study (data source: http://gaia.didichuxing.com). This study is supported by the General Research Fund #16212819 of the Research Grants Council of the HKSAR Government, and National Science Foundation of China (No. 71890970, No. 71890974). We thank the two anonymous reviewers and Associate Editor for their valuable comments.